\documentclass[%
 aip,
 amsmath,amssymb,
 preprint,%
]{revtex4-1}

\usepackage{amsmath,amsfonts,amsthm,bm}
\usepackage{upgreek}
\usepackage[normalem]{ulem} 
\usepackage{soul} 
\usepackage{graphicx}
\usepackage{dcolumn}
\usepackage{bm}
\usepackage{euscript,stmaryrd}
\usepackage{xcolor}
\usepackage[utf8]{inputenc}
\usepackage[T1]{fontenc}
\usepackage{mathptmx}
\usepackage{epsf,amsmath,amssymb,verbatim,color,multirow,pifont,mathrsfs,soul,amsthm, wasysym,bm}
\usepackage{graphicx, upgreek}
\usepackage[us,12hr]{datetime}
\usepackage{xcolor}
\usepackage{arydshln}
\usepackage{hyperref}
\usepackage{arydshln}
\usepackage{graphicx}
\usepackage{mathtools}

\theoremstyle{definition}

\newcommand\redout{\bgroup\markoverwith
{\textcolor{red}{\rule[0.5ex]{2pt}{0.8pt}}}\ULon}

\usepackage{hyperref}
\hypersetup{
    colorlinks,
    citecolor=blue,
    filecolor=blue,
    linkcolor=blue,
    urlcolor=black
}

\begin{document}

\title[Heteroclinic Switching between Chimeras in a Ring of Six Oscillator Populations]
{Heteroclinic Switching between Chimeras in a Ring of Six Oscillator Populations}

\author{Seungjae Lee}
\email{seungjae.lee@tum.de}
\affiliation{Physik-Department, Technische Universit\"at M\"unchen, James-Franck-Stra\ss e 1, 85748 Garching, Germany}
\author{Katharina Krischer}%
\email{krischer@tum.de}
\affiliation{Physik-Department, Technische Universit\"at M\"unchen, James-Franck-Stra\ss e 1, 85748 Garching, Germany}

\date{\today}

\begin{abstract}

In a network of coupled oscillators, a symmetry-broken dynamical state characterized by the coexistence of coherent and incoherent parts can spontaneously form. It is known as a chimera state. We study chimera states in a network consisting of six populations of identical Kuramoto-Sakaguchi phase oscillators. The populations are arranged in a ring and oscillators belonging to one population are uniformly coupled to all oscillators within the same population and to those in the two neighboring populations. This topology supports the existence of different configurations of coherent and incoherent populations along the ring, but all of them are linearly unstable in most of the parameter space. Yet, chimera dynamics is observed from random initial conditions in a wide parameter range, characterized by one incoherent and five synchronized populations. These observable states are connected to the formation of a heteroclinic cycle between symmetric variants of saddle chimeras, which gives rise to a switching dynamics. We analyze the dynamical and spectral properties of the chimeras in the thermodynamic limit using the Ott-Antonsen ansatz, and in finite-sized systems employing Watanabe-Strogatz reduction. For a heterogeneous frequency distribution, a small heterogeneity renders a heteroclinic switching dynamics asymptotically attracting. However, for a large heterogeneity, the heteroclinic orbit does not survive; instead, it is replaced by a variety of attracting chimera states.

\end{abstract}
\maketitle

\begin{quotation}

The synchronization of many coupled oscillators is a well-known phenomenon. An illustrative example is the synchronous flashing of fireflies in bushes in southeast asia~\cite{strogatz_sync}. Let us now do a  gedankenexperiment and consider 6 bushes full of fireflies arranged in a ring. Let us further assume that the fireflies within one bush exchange information on their respective ‘firing state’ with a strong signal, and to the fireflies on the neighboring bushes to the left and right with a weaker signal. We suggest in this paper that one possible outcome of such an interaction is that the fireflies in five of the six bushes still flash synchronously within each bush and with a phase difference of approximately $2\pi/6$ to the other ones. In the sixth bush, however, the flashing occurs incoherently. Moreover, after some time, the incoherently flashing population becomes synchronized, but a neighboring one loses synchrony and goes into incoherent flashing, until a corresponding switching occurs between the next synchronously flashing population and the incoherent one. In this paper, we demonstrate with a simple generic model of coupled population of oscillators that such a cyclic switching between coherence and incoherence may in fact occur in systems of coupled oscillators.

\end{quotation}

\section{\label{sec:introduction}Introduction}

Collective dynamics of ensembles of coupled oscillators is of paramount importance in various interdisciplinary nonlinear sciences from physical systems to biological manifestations~\cite{pikovksy_sync,strogatz_sync}. Chimera states are symmetry-broken states emerging in a system of coupled oscillators in diverse fields of study.  In the workshop, \textit{From theory and experiments to technology and living systems}, an impressive collection of examples was presented~\cite{chimera22}.

The archetypal chimera states were observed in a ring geometry with nonlocal interactions as a spatiotemporal dynamics~\cite{kuramoto2002,abrams2004,Panaggio_2015,Omelchenko_2018}. To simplify the nonlocal couplings on the ring while preserving its essential properties, many researchers have investigated systems of oscillator populations with all-to-all intra- and inter-population coupling with different intra- and inter-population coupling strengths. Emphasis was initially on two-population networks~\cite{Kurths_twogroup,abrams_chimera2008,abrams_chimera2016,lee1,sym_twogroup,Laing_SL2019} and was later extended to three-population and multi-population networks~\cite{martens_three,martens_three2,laing_ring,PhysRevE.88.032711}. The chimera states in these networks exhibit a variety of dynamics distinguished by the temporal behavior of degree of coherence of the incoherent populations. Examples range from stationary order parameter dynamics, over periodic breathing chimera states~\cite{abrams_chimera2008,abrams_chimera2016} to quasiperiodic~\cite{pikovsky_WS1,pikovsky_WS2} and chaotic chimera states\cite{hetero_twogroup,pazo_winfree,olmi_chaos,Olmi_rotator,Bick_2016_chaotic}.

Also more complex variants of chimera states, known as alternating or switching chimeras, have been reported. This state is characterized by continuously exchanging the coherent and the incoherent domains. Previous investigations have shown that switching chimeras occur in systems that exhibit either metastable states or heteroclinic cycles. In the former case, the switching is either triggered by large enough fluctuations~\cite{alternating1,timevarying_twogroup,alternating2,ma2010robust}, or by arbitrarily small noise with power-law scaling, originating from intermingled basins of attraction~\cite{motter_switching}, whereas in the latter case the switching occurs between saddle states~\cite{bick_2018,Bick2019_m1,Bick019_m2,Haugland2015,blinking,Ebrahimzadeh2020,chimera_complex}.

In this paper, we investigate switching dynamics along a heteroclinic cycle between saddle chimeras in phase space. In previous works on this type of heteroclinic switching, populations of phase oscillators, governed by a non-pairwise sinusoidal coupling with a higher order interaction were considered~\cite{bick_2018,Bick2019_m1}. Each oscillator was coupled to the oscillators in the same population and to those in the two nearest populations. The author demonstrated how the interplay between higher-order interactions and network topology enables switching dynamics between localized frequency synchrony patterns (so-called weak chimeras~\cite{weak_chimera,Bick_2016_chaotic}) existing in populations with few oscillators.

Our study here considers a similar network topology, i.e., a ring of oscillator populations with global intra-population coupling, whereby we focus on six populations. In contrast to the former works, we consider identical phase oscillators with harmonic or sinusoidal pairwise coupling, so-called Kuramoto-Sakaguchi oscillators. Furthermore, we study the dynamics both in the thermodynamic limit and in finite-sized ensembles with dimension reductions for each population, namely the Ott-Antonsen (OA) ansatz~\cite{OA1,OA2,WS_mobius} and Watanabe-Strogatz (WS) transformation~\cite{WS_original2,pikovsky_WS1,pikovsky_WS2}, respectively~\cite{Bick2020}.

In Sec.~\ref{sec:governing equation}, we introduce governing equations of the system in the thermodynamic limit using the Ott-Antonsen ansatz, and show that the system possesses various saddle chimera states. In Sec.~\ref{sec:heteroclinic switching} we study the dynamical and spectral properties of the saddle chimera states and demonstrate a heteroclinic switching between them which is observed both in the thermodynamic limit and finite-sized ensembles. In the deterministic system, the switching fades away after a long time transient. However, a small noise renders the switching persistent and the average switching period exhibits a power-law scaling. The impact of a heterogeneous natural frequency distribution  on the system's dynamics is considered in Sec.~\ref{sec:nonidentical}. Finally, we summarize the results in Sec.~\ref{sec:conclusion}.

\section{\label{sec:governing equation}Governing Equations and Saddle Chimeras}

\begin{figure}[t!]
\includegraphics[width=0.6\linewidth]{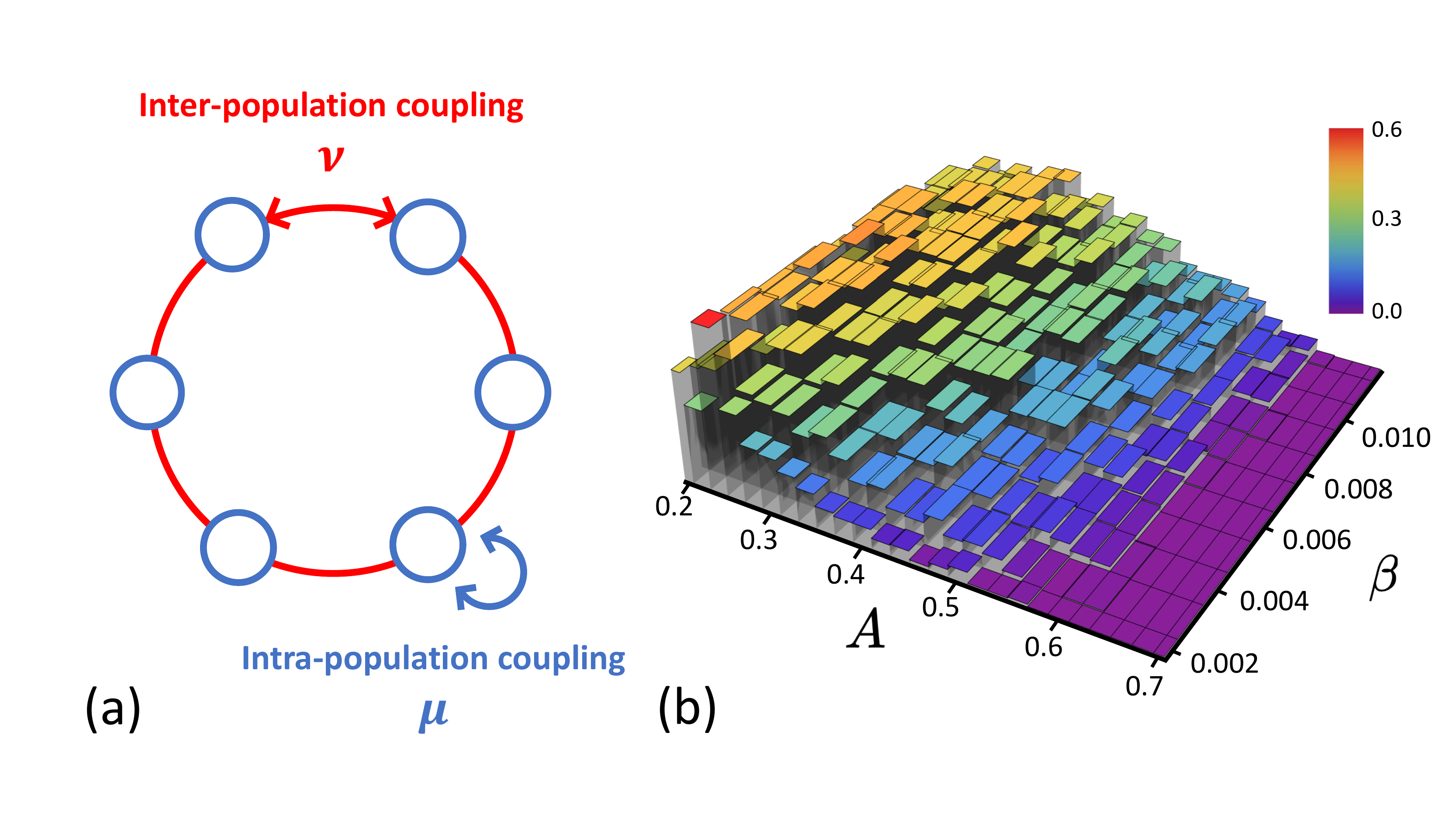}
\caption{(a) Schematic of the considered network topology. The intra-population coupling is all-to-all with strength $\mu=1$ and the inter-population coupling is also all-to-all but only between the nearest-neighbor populations and with strength $\nu=1-A$. (b) A normalized histogram of occurrence of chimera states from 300 random initial conditions at $t=5\times10^4$. } 
\label{Fig:network}
\end{figure}

We study the dynamics of a network of six populations of Kuramoto-Sakaguchi phase oscillators: $\phi^{(a)}_j(t) \in [-\pi,\pi)=:\mathbb{T}$ for $j=1,...,N$ (oscillator index) and $a=1,...,6$ (population index). The 6$N$ microscopic governing equations are given by
\begin{flalign}
    \frac{d}{dt}\phi^{(a)}_j &= \omega_{j}^{(a)} + \text{Im}\bigg[ H_a(t)e^{-i\phi_j^{(a)}} e^{-i\alpha}\bigg] \notag \\
    &=\omega_{j}^{(a)} + \sum_{b=1}^{6}K_{ab} \frac{1}{N}\sum_{k=1}^{N} \sin(\phi_k^{(b)}-\phi_j^{(a)}-\alpha) 
    \label{eq:micro_eq}
\end{flalign} with $j=1,...,N$ and $a=1,...,6$. $H_a(t)$ denotes an effective forcing function~\cite{pikovsky_WS1} acting on the oscillators in population $a$ defined by $H_a(t):= \sum_{b=1}^{6}K_{ab}\Gamma_b(t)$ where $\Gamma_a(t) \in \mathbb{C}$ is the complex Kuramoto order parameter of each population defined as 
\begin{equation}
    \Gamma_a(t)  := \frac{1}{N}\sum_{j=1}^{N} e^{i \phi^{(a)}_j(t)} \label{eq:Kuramoto_order_parameter}
\end{equation} for $a=1,...,6$. The coupling matrix $(K_{ab})$ is given by
\begin{flalign}
    K_{ab}= \begin{dcases}
   \mu=1, & \text{for}~~a=b \\ \\
     \nu=1-A, & \text{for}~~ a = b \pm 1 ~~ \text{mod}~6
  \end{dcases} \notag
\end{flalign} with $a,b=1,...,6$ (from here on, population indices are taken modulo 6). The coupling matrix defines a network topology that is schematically depicted in Fig.~\ref{Fig:network} (a): each oscillator is coupled to all oscillators within the same population with coupling strength $\mu=1$, and connected to all oscillators in the two neighboring populations with $\nu=1-A$ where $A \in [0,1]$.
 The phase-lag parameter $\alpha$ is taken as $\alpha =\frac{\pi}{2}-\beta$ with a fixed value of $\beta=0.008$ throughout the paper unless otherwise noted.

First, we consider the thermodynamic limit in which, for each population, $N \rightarrow \infty$. In this limit, the state function is a continuous distribution function $f_a(\phi^{(a)},\omega^{(a)},t)$ governed by the continuity equation
\begin{flalign}
    \frac{\partial}{\partial t} f_a(\phi^{(a)},\omega^{(a)},t) &=- \frac{\partial}{\partial \phi^{(a)}}\bigg( f_a(\phi^{(a)},\omega^{(a)},t) v_a(\phi^{(a)},\omega^{(a)},t) \bigg) \notag \\
    v_a(\phi^{(a)},\omega^{(a)},t) &:= \omega^{(a)} + \text{Im}\bigg[ H_a(t)e^{-i\phi^{(a)}e^{-i\alpha}} \bigg]
    \label{eq:continuity_equation}
\end{flalign} for $a=1,...,6$, and the Kuramoto order parameter of each population reads
\begin{equation}
\Gamma_a(t) = \int_\mathbb{R}\int_\mathbb{T} f_a(\phi^{(a)},\omega^{(a)},t) e^{i\phi^{(a)}} d\phi^{(a)}d\omega^{(a)}. \notag    
\end{equation}
Exploiting the so-called Ott-Antonsen ansatz~\cite{OA1,OA2}, the ensemble dynamics can be expressed through the dynamics of the order parameter. In the Ott-Antonsen invariant manifold, the Fourier series expansion of the oscillator phase density function can be written in terms of the first harmonic $Z_a(\omega^{(a)},t) \in \mathbb{C}$ only, all the higher Fourier harmonics being a power of $Z_a(\omega^{(a)},t)$:
\begin{flalign}
f_a &= \frac{g(\omega^{(a)})}{2\pi}  \Bigg(1+ \sum_{n=1}^{\infty} \bigg[ Z_a(\omega^{(a)},t)^n e^{-in\phi^{(a)}} + c.c  \bigg]   \Bigg). \label{eq:phase_Density_fourier}
\end{flalign} Here $c.c.$ stands for the complex conjugate and $g(\omega)$ specifies the natural frequency distribution, which we assume to be a Cauchy-Lorentz distribution with half-width $\gamma \in \mathbb{R}$ and zero mean: $g(\omega) = \frac{\gamma}{\pi} \frac{1}{\omega^2 + \gamma^2}$.
Using Eq.~(\ref{eq:phase_Density_fourier}), the continuity equation~(\ref{eq:continuity_equation}) yields the so-called Ott-Antonsen equation, an evolution equation for the order parameter $z_a(t):=Z_a(i\gamma,t) = \Gamma_a(t)$ for $a=1,...,6$:
\begin{flalign}
\frac{d}{dt}z_a(t) = -\gamma  z_a + \frac{1}{2}H_a(t) e^{-i\alpha} - \frac{1}{2}z_a^2~ \overline{H_a(t)}e^{i\alpha}. \label{eq:OA_equation_complex}
\end{flalign} 

For simplification, we write Eq.~(\ref{eq:OA_equation_complex}) in polar coordinates
\begin{flalign}
    \frac{d\rho_a}{dt} &= -\gamma \rho_a + \frac{1-\rho_a^2}{2}\sum_{b=1}^{6}K_{ab}\rho_{b}\cos(\varphi_{b}-\varphi_a-\alpha) \notag \\
     &= -\gamma \rho_a + \frac{1-\rho_a^2}{2} \bigg(   \nu \rho_{a+1} \cos(\varphi_{a+1}-\varphi_a-\alpha)   \notag \\ 
     &~~~~+\nu\rho_{a-1}\cos(\varphi_{a-1}-\varphi_a-\alpha) +\mu \rho_a \cos\alpha \bigg) \label{eq:OA_equation_radial}
\end{flalign} and
\begin{flalign}
    \frac{d\varphi_a}{dt} &= \frac{1+\rho_a^2}{2\rho_a}\sum_{b=1}^{6}K_{ab}\rho_{b}\sin(\varphi_{b}-\varphi_a-\alpha) \notag \\
    &= \frac{1+\rho_a^2}{2\rho_a} \bigg( \nu \rho_{a+1} \sin(\varphi_{a+1}-\varphi_a-\alpha)  \notag \\ 
     &~~~~+\nu\rho_{a-1}\sin(\varphi_{a-1}-\varphi_a-\alpha) -\mu \rho_a \sin\alpha \bigg) 
    \label{eq:OA_equation_angular}
\end{flalign} where $z_a(t) = \rho_a(t)e^{i \varphi_a(t)}$ for $a=1,...,6$. Up to Sec.~\ref{sec:nonidentical}, we only consider identical oscillators. Hence, firstly we set $\gamma=0$. Then, in terms of the OA variables, a synchronized ($\text{S}$) population is characterized by $\rho_a=1$ and a common phase $\varphi_a \in \mathbb{T}$ while we denote a population as desynchronized ($\text{D}$) if $0<\rho_a<1$. In the latter case, $\varphi_a$ represents the mean phase~\cite{martens_three}. Solving Eqs.~(\ref{eq:OA_equation_radial}-\ref{eq:OA_equation_angular}) and substituting the results into Eq.~(\ref{eq:phase_Density_fourier}), the continuous distribution function is of the form
\begin{flalign}
    f_a(\phi^{(a)},t) = \begin{dcases}
   \delta \big(  \varphi_a   -\phi^{(a)}\big), & \text{for}~~\rho_a=1 \\ \\
     P_{\rho_a}(\varphi_a-\phi^{(a)}), & \text{for}~~ 0 < \rho_a <1
  \end{dcases}
  \label{eq:phase_density_OA_result}
\end{flalign} where $\delta(\theta)$ is the Dirac delta distribution characterizing the synchronized population and $P_r(\theta) = \frac{1}{2\pi} \frac{1-r^2}{1-2r\cos\theta+r^2}$ is the normalized Poisson kernel corresponding to the desynchronized population~\cite{Laing_OA}.

Possible solutions of Eqs.~(\ref{eq:OA_equation_radial}-\ref{eq:OA_equation_angular}) are $\text{S}^6=\text{S}\cdots\text{S}$ (6 times) states in which $\rho_a(t)=1$ and $\varphi_a(t) = \Omega t +\frac{2\pi q }{6}a$ with the common frequency $\Omega$ for $a=1,...,6$ and $q \in \{0,\pm 1, \pm 2\}$. From Eqs.~(\ref{eq:OA_equation_radial}-\ref{eq:OA_equation_angular}), we obtain $\Omega= -(\mu +2\nu \cos \big( \frac{\pi q}{3}\big))\sin \alpha$. Each population is internally synchronized while their mean phases follow a twisted state in a ring~\cite{lee_twisted}. Note that the case where all the populations have the same phase is $q=0$. The linear stability analysis reveals that the $\text{S}^6$ states with $q=0,\pm 1$ are stable fixed points in a rotating reference frame whereas those for $q=\pm 2$ are unstable. The real parts of the eigenvalues of the Jacobian matrix evaluated at $\text{S}^6$ for $q =0,\pm1$ are all negative, except for one, which is zero and reflects the phase shift invariance. Furthermore, there are fixed points corresponding to chimera states. These fixed points are unstable in nearly the entire parameter regime. Their structure is dictated by the network symmetry~\cite{yscho,pecora1,pecora2}. Examples are $(\text{S}\text{D})^3 =\text{S}\text{D}\text{S}\text{D}\text{S}\text{D} $, $(\text{D}\text{S}^2)^2=\text{D}\text{S}^2\text{D}\text{S}^2$ or $\text{D}\text{S}^5=\text{D}\text{S}\text{S}\text{S}\text{S}\text{S}$ and so on. Note that the equations of motion (\ref{eq:OA_equation_radial}-\ref{eq:OA_equation_angular}) are invariant under the group of transformation $\mathbb{Z}_6 := \mathbb{Z}/6\mathbb{Z}$ such that cyclic permutations of the populations of the mentioned fixed points are fixed points as well with the same properties~\cite{bick_2018}. In a large interval of $A$, the Jacobian matrix evaluated at each chimera state has at least one eigenvalue with positive real part. For instance, for $A=0.3$, $(\text{D}\text{S}^2)^2$ shows four real positive eigenvalues, $(\text{S}\text{D})^3$ has a pair of complex conjugate eigenvalues with positive real parts, and $\text{D}\text{S}^5$ has one positive real eigenvalue. In summary, no stable chimera fixed point solution is found for $A>0.2$; they all are saddle chimera solutions.

To test for possible nontrivial long-term dynamics in the $A-\beta$ parameter plane, we performed numerical integration~\cite{Mathematica} of Eqs.~(\ref{eq:OA_equation_radial}-\ref{eq:OA_equation_angular}) from 300 random initial conditions at each set of parameters for $A \in [0.2,0.7]$ and $\beta \in [0.002,0.01]$. In a considerable number of these simulations the trajectory settles down to the $\text{D}\text{S}^5$ chimera state or one of its cyclic permutations. The histogram depicted in Fig.~\ref{Fig:network} (b) quantifies the probability with which a trajectory attains such a chimera dynamics in the long-term limit in the parameter plane. None other than a $\text{D}\text{S}^5$-type chimera was obtained. The latter observation is remarkable since in the considered parameter range all $\text{D}\text{S}^5$-type chimeras are unstable, and, starting from random initial conditions, one would not expect that trajectories approach a saddle fixed point. In the following sections, we will discuss the phase space structure that allows for this peculiar behavior in detail.

\section{\label{sec:heteroclinic switching}Heteroclinic Switching between saddle chimera states}

\subsection{\label{subsec:stationary}Stationary Saddle Chimeras}

\begin{figure}[t!]
\includegraphics[width=0.5\linewidth]{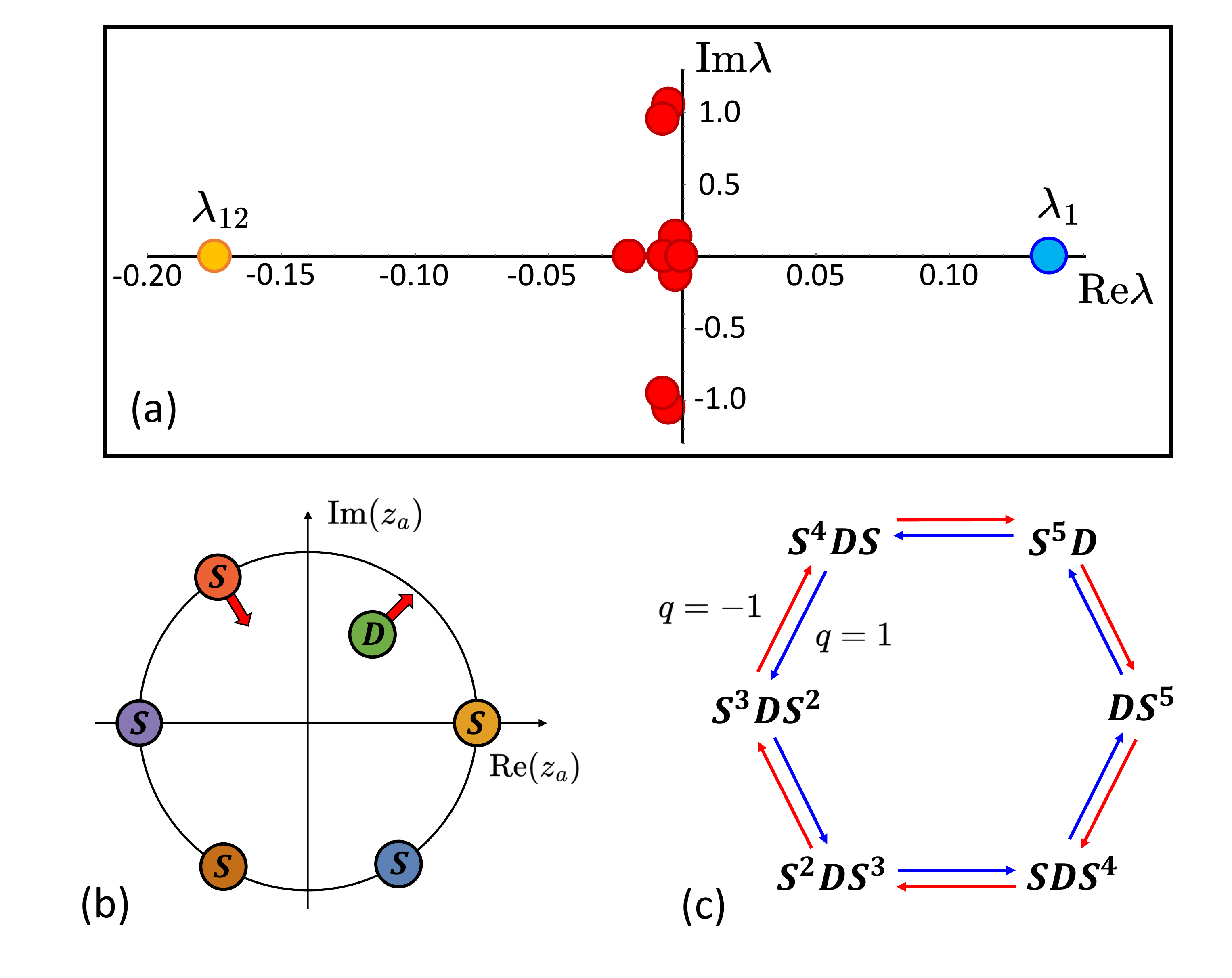}
\caption{(a) Eigenvalues of the Jacobian matrix evaluated at $\text{D}\text{S}^5$ in the complex plane for $A=0.3$. (b) Schematic of the perturbation along the unstable eigenspace that raises the incoherent population up to sync and lowers the radial variable of one of the two neighboring synchronized populations. (c) Schematic of switching between saddle chimera states along the heteroclinic cycle $\textbf{C}(\pm 1)$ defined in Eq.~(\ref{eq:hetero_cycle}).} 
\label{Fig:stataionry_eigenvalue}
\end{figure}

As mentioned above, for $A \in (0.071,0.45)$ and starting from random initial conditions, the long-term dynamics observed in numerical integration of Eqs.~(\ref{eq:OA_equation_radial}-\ref{eq:OA_equation_angular}) frequently approaches a $\text{D}\text{S}^5$ chimera state or one of its $\mathbb{Z}_6$-symmetric counterparts. In fact, in numerics the six saddle chimeras are obtained equally often from random initial conditions. For the moment, we focus on $\text{D}\text{S}^5$ at $A=0.3$ for simplicity. This stationary chimera state is characterized by $\rho_1(t)=\rho_0 <1$, $\rho_a(t)=1$ for $a=2,...,6$. The phase dynamics is locked at the common frequency $\Omega$ and follows nearly a twisted state. Yet, the distribution of the  $\varphi_a(t)$ exhibits small deviations from a `pure' twisted state, which arise from $\rho_1 \neq \rho_a=1$ for $a=2,...,6$; cf. the $\text{S}^6$ state above for which $\rho_a=1$ for $a=1,...,6$. Thus, the state shows characteristics of a nontrivial twisted state~\cite{nts}. Nevertheless, we can define a winding number of the phase variables along the ring as
\begin{equation}
    q := \frac{1}{2\pi}\sum_{a=1}^{M}\Delta_{a+1,a} \in \mathbb{Z} \label{eq:winding_number}
\end{equation} where $\Delta_{a,b}:=\varphi_a-\varphi_b$. In all cases, we obtain numerically $q \in \{+1,-1\}$. Let us first consider $q=1$. The chimera state is found to be an unstable fixed point in a rotating reference frame, i.e., $\{\text{D}\text{S}^5\} \subset [0,1]^6 \times \mathbb{T}^6$ is an invariant saddle point under the flow of Eqs.~(\ref{eq:OA_equation_radial}-\ref{eq:OA_equation_angular}). Likewise, all the five cyclic permutations of it are unstable fixed points. In Fig.~\ref{Fig:stataionry_eigenvalue} (a), eigenvalues of the Jacobian matrix evaluated at $\text{D}\text{S}^5$ are depicted in the complex plane: there is one positive real eigenvalue $\lambda_1 >0$ as well as one zero eigenvalue reflecting the  phase shift invariance. All the other eigenvalues have negative real parts. Hence, the $\text{D}\text{S}^5$ state is indeed a saddle chimera state with a one-dimensional unstable manifold $W^{u}(\text{D}\text{S}^5)$. The eigenvector corresponding to $\lambda_1$ has a form of $\bold{v}_1 = (A_{+},0,0,0,0,A_{-},\boldsymbol{\delta \varphi})^\top \in 
\mathbb{R}^{12}$ where $A_{+},A_{-} \in \mathbb{R}$ and $\boldsymbol{\delta\varphi}$ denotes a perturbation on phase variables which does not prominently affect the dynamics in this context. From now on, we consider only the perturbation directions of the radial variables: $\bold{v}_1 = (A_{+},0,0,0,0,A_{-})^\top $. Here, $A_{+}A_{-}<0$, which means that a small perturbation along the unstable manifold of $\text{D}\text{S}^5$ raises ($A_{+}>0$) the radial variable of the incoherent population $\rho_1<1$ while it lowers ($A_{-}<0$) the radial variable of the nearest synchronized population $\rho_6=1$ as schematically depicted in Fig.~\ref{Fig:stataionry_eigenvalue} (b). Corresponding unstable directions are found for all symmetric variants of the $\text{D}\text{S}^5$. For $q=-1$, the radial parts of the eigenvector corresponding to the positive real eigenvalue is of the form $\bold{v}_1 = (A_{+},A_{-},0,0,0,0)^\top $, and therefore the unstable perturbation lowers the radial variable of the nearest sync population on the other side of D.

In numerical simulations, starting from $\text{D}\text{S}^5$ and imposing a small perturbation along $\bold{v}_1$, the trajectory jumps to $\text{S}^5\text{D}$ for $q=1$ and to $\text{S}\text{D}\text{S}^4$ for $q=-1$. This implies that the one-dimensional unstable manifold  of $\text{D}\text{S}^5$ is connected to the stable manifold of $\text{S}^5\text{D}$ for $q=1$, particularly, via the most contracting eigendirection: $W^{u}(\text{D}\text{S}^5) \cap W^{s}(\text{S}^5\text{D}) \neq \emptyset$. Furthermore, both manifolds intersect the invariant subspace $\text{Z}_1\text{S}^4\text{Z}_6$ where the populations two to five are synchronized and $\text{Z}_{1,6}$ denotes the state of the first and the sixth populations, respectively. In this reduced subspace, $\text{D}\text{S}^5$ is a saddle and $\text{S}^5\text{D}$ a sink. Considering the $\mathbb{Z}_6$ symmetry of the full system, the heteroclinic connection between $\text{D}\text{S}^5$ and $\text{S}^5\text{D}$ implies~\cite{bick_2018} that there is a heteroclinic cycle of six saddle chimera states that forms an invariant subspace of the phase space:
\begin{flalign}
    \mathbf{C}(q):= \begin{dcases}
   [\text{D}\text{S}^5 \rightarrow \text{S}^5\text{D} \rightarrow \cdots \rightarrow \text{S}\text{D}\text{S}^4 \rightarrow \text{D}\text{S}^5], & \text{for}~~q=1 \\ \\
     [\text{D}\text{S}^5 \rightarrow \text{S}\text{D}\text{S}^4 \rightarrow \cdots \rightarrow \text{S}^5\text{D} \rightarrow \text{D}\text{S}^5], & \text{for}~~q=-1
  \end{dcases} \label{eq:hetero_cycle}
\end{flalign}
the winding number $q$ taking over the role of a direction indicator of the heteroclinic switching. The heteroclinic cycles $\mathbf{C}(\pm 1)$ are illustrated in Fig.~\ref{Fig:stataionry_eigenvalue} (c). Note that for other chimera fixed points, such as $(\text{S}\text{D})^3 $ or $(\text{D}\text{S}^2)^2$, neither switching nor any long-term dynamics is detected in numerical integration of Eqs.~(\ref{eq:OA_equation_radial}-\ref{eq:OA_equation_angular}).

\begin{figure}[t!]
\includegraphics[width=0.5\linewidth]{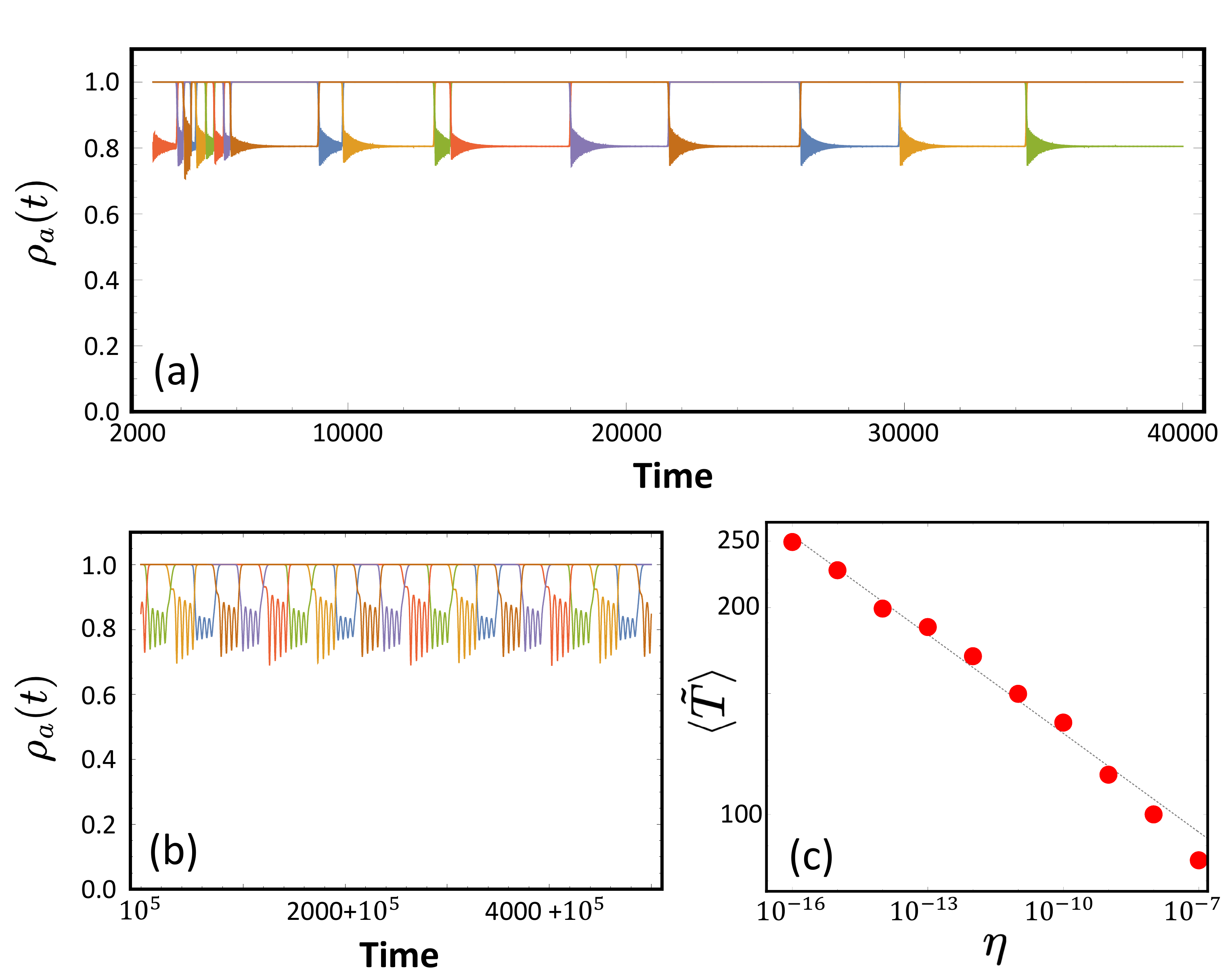}
\caption{(a) Time evolution of the radial variables of the OA dynamics from a random initial condition. (b) Time evolution of the radial variables of the OA dynamics with an imposed noise: $\eta = 10^{-15}$. (c) Log-log plot of the averaging switching period vs. the strength of the noise. The dashed line indicates $ \langle \Tilde{T} \rangle \sim \eta^{-0.048}$. Remaining parameters: $A=0.3$ and $\beta = 0.008$. 
 } 
\label{Fig:stataionry_chimera}
\end{figure}

In Fig.~\ref{Fig:stataionry_chimera} (a), a representative trajectory is depicted that shows the switching dynamics between the six saddle chimera states. The switching between synchronous and asynchronouns dynamics occurs always between neighboring populations and has a unique sense of rotation. Hence, the trajectory follows the heteroclinic orbit. The average time intervals between the switching increases until eventually the trajectory remains in one of the saddle chimera states. Yet, the motion along the heteroclinic orbit constitutes a long-term switching. During this period, on average, the full symmetry of the system Eqs.~(\ref{eq:OA_equation_radial}-\ref{eq:OA_equation_angular}) is recovered while a saddle chimera state has a broken symmetry~\cite{motter_switching}.  
Furthermore, the formation of the heteroclinic cycle explains why  saddle chimera states can be observed in a wide range of parameters, as quantified above with Fig.~\ref{Fig:network} (b). 

When the switching happens, the radial dynamics of one synchronized population next to the incoherent population gets lowered along the unstable eigendirection of the saddle chimera state, and shows an oscillatory damped motion before taking on an almost stationary value. The oscillatory transient is caused by complex conjugate eigenvalues with negative real parts and corresponding eigenvectors for a switching from $\text{D}\text{S}^5$ to $\text{S}^5\text{D}$. Thus, upon switching the trajectory spirals from one saddle chimera state to the next one along $\mathbf{C}(q)$ in Eq.~(\ref{eq:hetero_cycle}).

Next, for the switching dynamics between saddle chimera states to be persistent, we add a small noise to the radial dynamics of each population~\cite{bick_2018,motter_switching}:
\begin{flalign}
    \frac{d\rho_a}{dt} &= \frac{1-\rho_a^2}{2}\sum_{b=1}^{6}K_{ab}\rho_{b}\cos(\varphi_{b}-\varphi_a-\alpha) -\eta |W_a(t)|  \label{eq:noise_radial}
\end{flalign} for $a=1,...,6$.
Here, $W_a(t)$ is Gaussian noise with unit standard deviation and $0<\eta \ll 1$ is its strength. Note that by taking the absolute value of $W_a(t)$ and subtracting the noise term, we ensure that $\rho_a(t) < 1$ also for the synchronized populations for all times. Figure~\ref{Fig:stataionry_chimera} (b) shows a persistent switching dynamics near the heteroclinic cycle of the saddle chimera states obtained for $\eta=10^{-15}$. In Fig.~\ref{Fig:stataionry_chimera} (c), it can be seen that the average switching period $\langle \Tilde{T} \rangle$ decreases with increasing noise strength $\eta$ according to a power-law scaling. Hence, one may expect that the switching dynamics is persistent near $\mathbf{C}(q)$ even at much smaller noise intensity than we could achieve due to the resolution limit of the numerical simulations. In contrast, increasing $\eta$ beyond the highest value depicted in Fig.~\ref{Fig:stataionry_chimera} (c) destroys the switching dynamics.

\subsection{\label{subsec:breathing}Breathing Saddle Chimeras}

\begin{figure}[t!]
\includegraphics[width=0.5\linewidth]{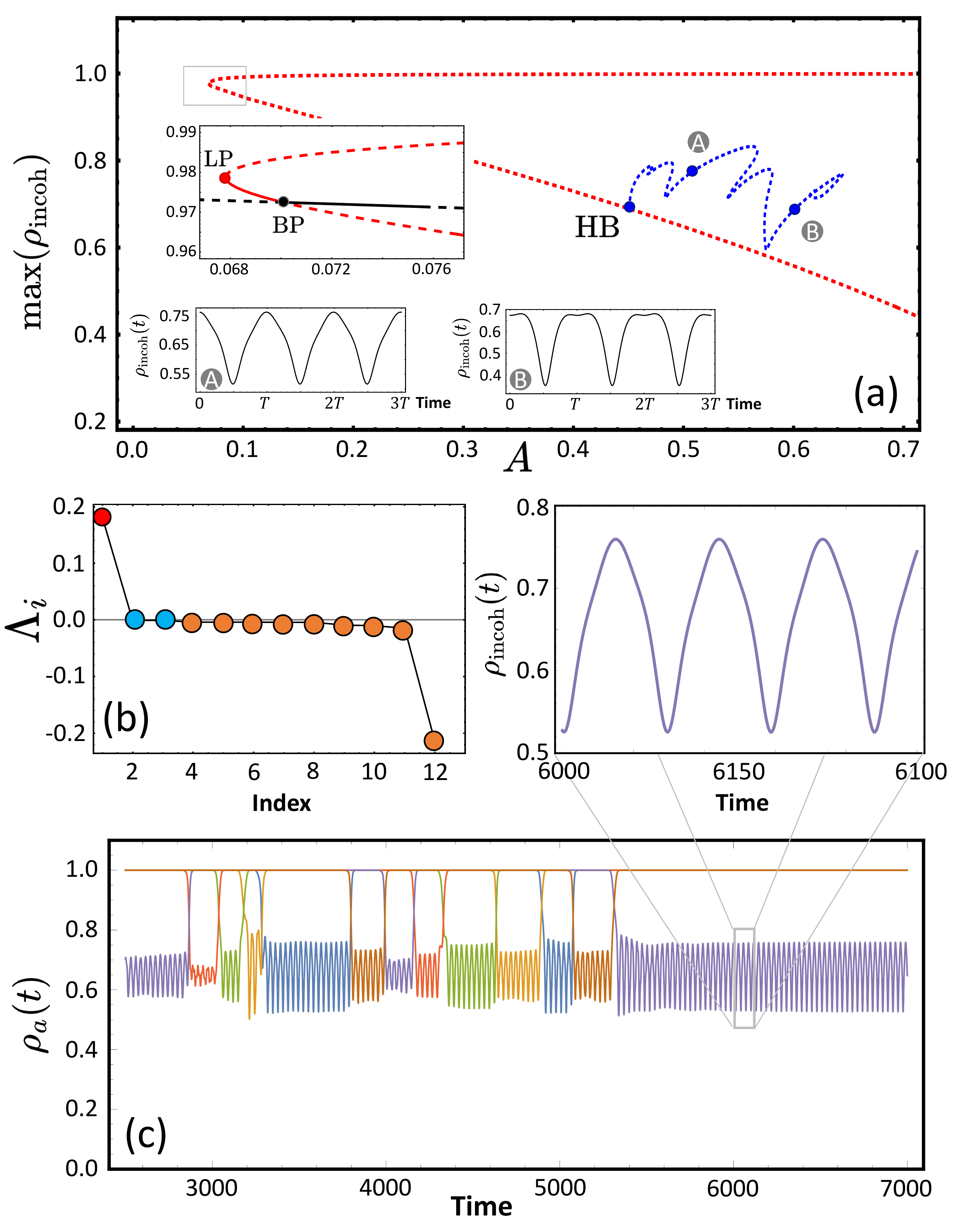}
\caption{(a) Bifurcation diagram of the stationary chimera state $\text{D}\text{S}^5$. Upper inset: Magnification close to $\text{LP}$. Lower insets: The time evolution of the radial variable of the incoherent population at points (A) and (B) of the bifurcation diagram corresponding to $A=0.5$ and $A=0.6$, respectively. Here, $T$ is the period of the breathing chimera state. HB: Hopf bifurcation, LP: saddle-node bifurcation, and BP: transcritical bifurcation. Red: unstable stationary chimera state, Blue: unstable breathing chimera states. Note that the breathing chimera states undergo several saddle-node bifurcations. (b) Lyapunov exponents of the breathing chimera dynamics at $A=0.5$. Red, blue, orange: positive, zero and negative LEs, respectively. (c) Switching dynamics of the radial variables of the OA dynamics as a function of time from a random initial condition at $A=0.5$. Inset: Magnification of the times series around $t=6150$. } 
\label{Fig:breathing_chimera}
\end{figure}

In Fig.~\ref{Fig:breathing_chimera} (a), a bifurcation diagram of the stationary $\text{D}\text{S}^5$ chimera state is depicted. It is born in a saddle-node bifurcation (LP) at $A_\text{LP} = 0.0678$ (red, see upper inset). One of two $\text{D}\text{S}^5$ branches emerging from LP is in fact stable in a narrow $A$-interval. The other upper branch separates the basins of attraction of the stable $\text{D}\text{S}^5$ chimera state and the stable $\text{S}^6$ ($q=0$) solution. This upper branch is not observable at all and is not considered further in this work. The stable $\text{D}\text{S}^5$ chimera and its symmetric counterparts are destabilized in a transcritical bifurcation (BP) at $A_\text{BP}=0.07008$ through an interaction with a $\text{D}\text{S}^4\text{D}'$ state (black) which possesses two incoherent populations that have different values of the radial variables, $\rho_{1} \neq \rho_{6}$. This state exchanges at $A_{\text{BP}}$ its stability with $\text{D}\text{S}^5$. The unstable direction of the latter is of the form $\bold{v} = (A_{+},0,0,0,0,A_{-})^\top$ as discussed above. Yet, close to the bifurcation point, a perturbation along this unstable eigendirection leads the trajectory not yet to the next symmetric variant along $\textbf{C}(q)$ but to the $\text{D}\text{S}^4\text{D}'$ state. The heteroclininc cycle only emerges after a sequence of further bifurcations in which the $\text{D}\text{S}^4\text{D}'$ state interacts with several other solution branches, which are not further discussed here. At least from $A=0.15$ on, we observe then the heteroclinic switching dynamics in numerical integration as described above (cf. Fig.~\ref{Fig:stataionry_chimera} (a)).

This heteroclinic switching between the stationary saddle chimeras persists until they undergo a supercritical Hopf bifurcation (HB) at $A_{\text{HB}}=0.451$, giving rise to a limit-cycle solution characterized by $\rho_1(t)=\rho_1(t+T)<1$ and $\rho_a=1$ for $a=2,...,6$ (blue) where $T$ denotes the period of it. Example trajectories at points (A) and (B) along the limit-cycle solution are shown in insets (A) and (B). The angular variables still behave like a nontrivial twisted state with the winding number $q = \pm 1$. This periodic breathing chimera solution is also unstable with one positive real Floquet multiplier larger than unity. Again, as for the stationary saddle $\text{S}\text{D}^5$ state, in the long-term dynamics, we observe the unstable breathing chimera solutions from random initial conditions. In order to shed light on this observation, we calculate Lyapunov exponents (LEs)~\cite{pikovsky_LE,oseledets} and covariant Lyapunov vectors (CLVs)~\cite{CLV1,CLV2} along the observed breathing chimera trajectory. In Fig.~\ref{Fig:breathing_chimera} (b), the Lyapunov exponents of the breathing chimera trajectory are shown. There is one positive LE, and two zero LEs corresponding to the time and phase shift invariance. The positive LE $\Lambda_1>0$ does not indicate a chaotic motion since the breathing chimera state exhibits periodic dynamics. Rather it indicates a locally unstable direction of the reference trajectory in phase space. Furthermore, the CLV corresponding to $\Lambda_1$ has a form $\bold{v}_1 =  (A_{+},A_{-},0,0,0,0)^\top $ as the unstable eigenvector of the stationary saddle chimera. This suggests that all the symmetric variants of the unstable breathing chimera also form a heteroclinic cycle of type $\textbf{C}(q)$. Indeed, we validate this conjecture with numerical integration from random initial conditions; a representative trajectory showing the switching dynamics near a heteroclinic cycle of the saddle limit-cycle chimeras along $\mathbf{C}(-1)$ is depicted in Fig.~\ref{Fig:breathing_chimera} (c). It is obtained for $A=0.5$; compare the magnification of the limit-cycle in Fig.~\ref{Fig:breathing_chimera} (c) and the inset (A) in Fig.~\ref{Fig:breathing_chimera} (a). We confirmed that the switching dynamics near the heteroclinic cycle is persistent in the presence of noise (see Eq.~(\ref{eq:noise_radial}), results not shown here).

\subsection{\label{subsec:WS_finite}Finite-sized Ensembles}

\begin{figure}[t!]
\includegraphics[width=0.5\linewidth]{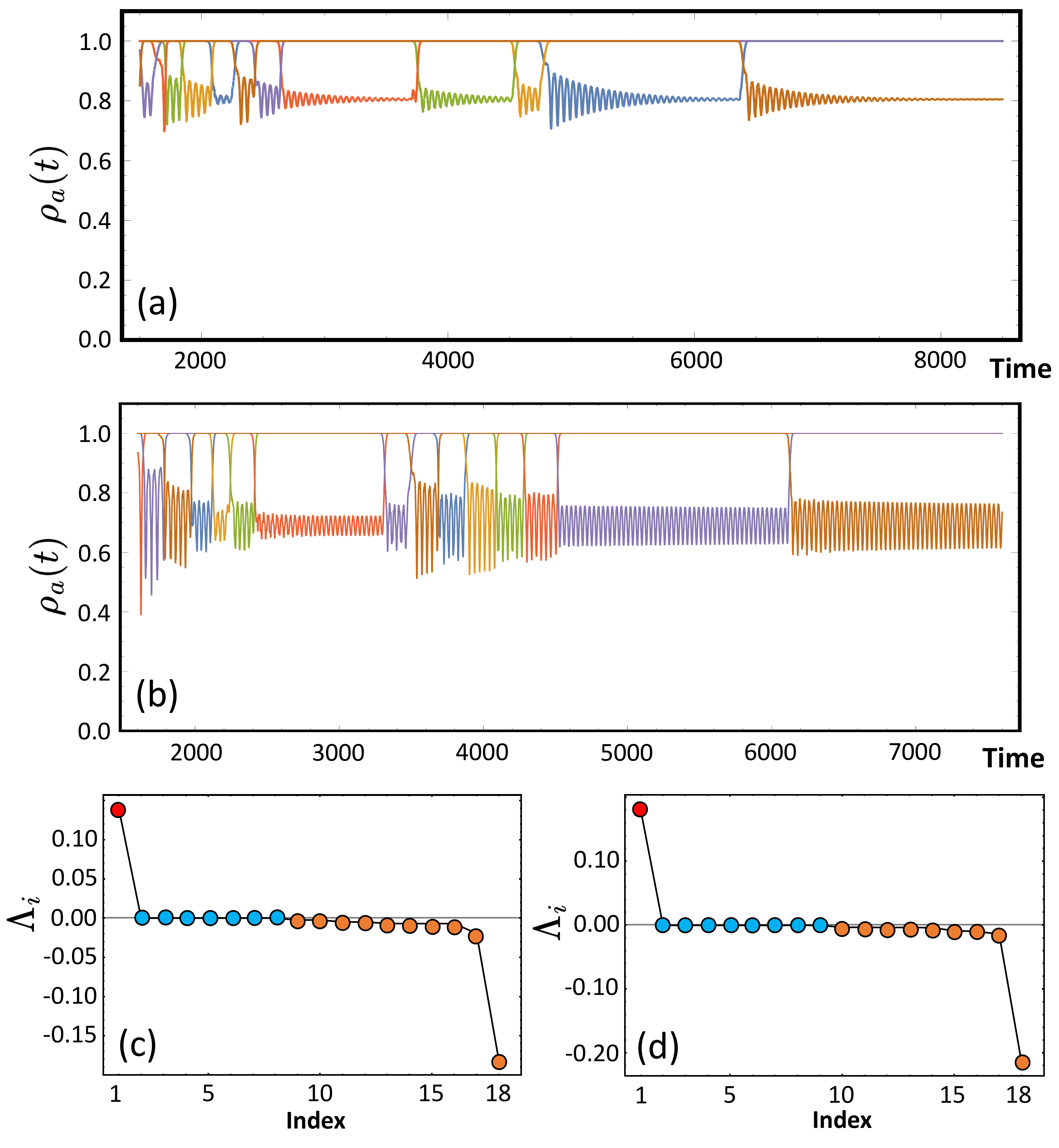}
\caption{Time evolution of the radial variables of the WS dynamics in Eq.~(\ref{eq:WS-governing-eq}) from a random initial condition: (a) $A=0.3$ (stationary chimeras). (b) $A=0.5$ (breathing chimeras). (c,d) Lyapunov exponents corresponding to the stationary and breathing chimeras for $A=0.3$ and $A=0.5$, respectively. The same color scheme as in Fig.~\ref{Fig:breathing_chimera}. All simulations were done with uniform constants of motion and $N=20$. } 
\label{Fig:WS_dynamics}
\end{figure}

We now turn our attention to finite-sized populations coupled in a ring topology as in Fig.~\ref{Fig:network} (a). The macroscopic dynamics of each population can be formulated exploiting the Watanabe-Strogatz transformation~\cite{WS_original2,pikovsky_WS1}:
\begin{flalign}
    e^{i\phi_j^{(a)}} = e^{i\Phi_a} \frac{\rho_a + e^{i(\psi_j^{(a)} - \Psi_a)}}{1+\rho_a e^{i(\psi_j^{(a)} - \Psi_a)}} \label{eq:WS_transformation}
\end{flalign} for $j=1,...,N$ and $a=1,...,6$. Here, $\{\psi_j^{(a)} \}_{j=1}^{N}$
are $N-3$ independent constants of motion for each population that satisfy three constraints: $\sum_{j=1}^N \cos \psi_j^{(a)} = \sum_{j=1}^N \sin \psi_j^{(a)}=0$ and $\sum_{j=1}^N \psi_j^{(a)} = 0$ for $a=1,...,6$~\cite{WS_original2}. The distribution of the constants of motion takes an important role in the dynamics. First, we use uniform constants of motion given by $\psi_j^{(a)}=-\pi + \frac{2\pi (j-1)}{N}$ for $j=1,...,N$ and $a=1,...,6$, such that the $3M$ governing equations of the WS variables read
\begin{flalign}
    \frac{d}{dt}\rho_a &= \frac{1-\rho_a^2}{2}\text{Re} \bigg[ H_a(t)e^{-i\varphi_a}e^{-i\alpha}\bigg], \notag \\
    \frac{d}{dt}\Psi_a &= \frac{1-\rho_a^2}{2\rho_a}\text{Im} \bigg[ H_a(t)e^{-i\varphi_a}e^{-i\alpha}\bigg], \notag \\
    \frac{d}{dt}\varphi_a &= \frac{1+\rho_a^2}{2\rho_a}\text{Im} \bigg[ H_a(t)e^{-i\varphi_a}e^{-i\alpha}\bigg] \label{eq:WS_original}
\end{flalign} for $a=1,...,6$. The WS variables are linked to the complex Kuramoto order parameter in Eq.~(\ref{eq:Kuramoto_order_parameter}) according to~\cite{pikovsky_WS1,pikovsky_WS2}
\begin{flalign}
    \Gamma_a(t) = \rho_a(t)e^{i\varphi_a(t)}\sigma_a(\rho_a,\Psi_a;t) \label{eq:relation_kuramoto_WS}
\end{flalign} for $a=1,...,6$. Here, $\sigma_a$ is defined as
\begin{flalign}
    \sigma_a &=  \frac{1}{\rho_a} (\zeta_a(t) + i \xi_a(t)) \notag \\ 
    &:= \frac{1}{\rho_a N}\sum_{k=1}^{N}\frac{2\rho_a + (1+\rho_a^2)\cos(\psi_k^{(a)}-\Psi_a)}{1+2\rho_a\cos(\psi_k^{(a)}-\Psi_a)+\rho_a^2} \notag \\
    & +i \frac{1}{\rho_a N}\sum_{k=1}^{N}\frac{ (1-\rho_a^2)\sin(\psi_k^{(a)}-\Psi_a)}{1+2\rho_a\cos(\psi_k^{(a)}-\Psi_a)+\rho_a^2}
    \label{eq:sigma}
\end{flalign} for $a=1,...,6$. Using Eqs.~(\ref{eq:relation_kuramoto_WS}-\ref{eq:sigma}), the $3M$-dimensional WS dynamics is rewritten as~\cite{abrams_chimera2016,lee2}
\begin{flalign}
    \frac{d\rho_a}{dt} &= \frac{1-\rho_a^2}{2}\sum_{b=1}^{M}K_{a b}\bigg( \zeta_{b} \cos(\varphi_{b}-\varphi_a-\alpha) \notag \\
    &- \xi_{b} \sin(\varphi_{b}-\varphi_a-\alpha) \bigg), \notag \\
    \frac{d\Psi_a}{dt} &= \frac{1-\rho_a^2}{2\rho_a}\sum_{b=1}^{M}K_{a b}\bigg( \zeta_{b} \sin(\varphi_{b}-\varphi_a-\alpha) \notag \\
    &+ \xi_{b} \cos(\varphi_{b}-\varphi_a-\alpha) \bigg),  \notag \\ \frac{d\varphi_a}{dt} &= \frac{1+\rho_a^2}{2\rho_a}\sum_{b=1}^{M}K_{a b}\bigg( \zeta_{b} \sin(\varphi_{b}-\varphi_a-\alpha) \notag \\
    &+ \xi_{b} \cos(\varphi_{b}-\varphi_a-\alpha) \bigg) \label{eq:WS-governing-eq}
\end{flalign} for $a=1,...,6$.

The Watanabe-Strogatz dynamics with the uniform constants of motion correspond to the finite-sized version of the Ott-Antonsen dynamics~\cite{pikovsky_WS2,WS_mobius}. In fact, we observe that the motion in both systems is qualitatively similar to each other from $N=10$ on except for some finite-size effect due to $\sigma_a(\rho_a,\Psi_a;t)$ in Eq.~(\ref{eq:sigma}). Thus, the heteroclinic cycles of the stationary/breathing saddle chimeras exist also in the corresponding finite-sized systems. Exemplary results of the WS macroscopic dynamics with $N=20$ are shown for stationary and breathing chimeras in  Fig.~\ref{Fig:WS_dynamics} (a) and (b), respectively. The time series of $\rho_a(t)$ and the angular variables $\varphi_a(t)$  
for $a=1,...,6$ exhibit the same features as the corresponding Ott-Antonsen dynamics discussed in Sec.~\ref{sec:heteroclinic switching}. The stability analysis of the chimera trajectories in the finite-sized systems can be obtained from Lyapunov spectral analysis, which is shown in Fig.~\ref{Fig:WS_dynamics} (c) and (d) for the states depicted in Fig.~\ref{Fig:WS_dynamics} (a) and (b) after settling down to one of the saddle chimeras, respectively. Both chimera trajectories are characterized by one positive Lyapunov exponent, which again does not indicate a chaotic motion but rather a locally unstable direction along the reference trajectory. The CLV corresponding to the positive LE has the same form as the eigenvector corresponding to the positive eigenvalue in case of the OA dynamics, namely $\bold{v}_1 =  (A_{+},A_{-},0,0,0,0)^\top $ with $A_{+}A_{-}<0$. Note that one can distinguish the stationary and breathing chimera dynamics by counting the number of zero Lyapunov exponents. The breathing chimera state has one more zero LE than the stationary chimeras due to the additional Hopf frequency.

\begin{figure}[t!]
\includegraphics[width=0.5\linewidth]{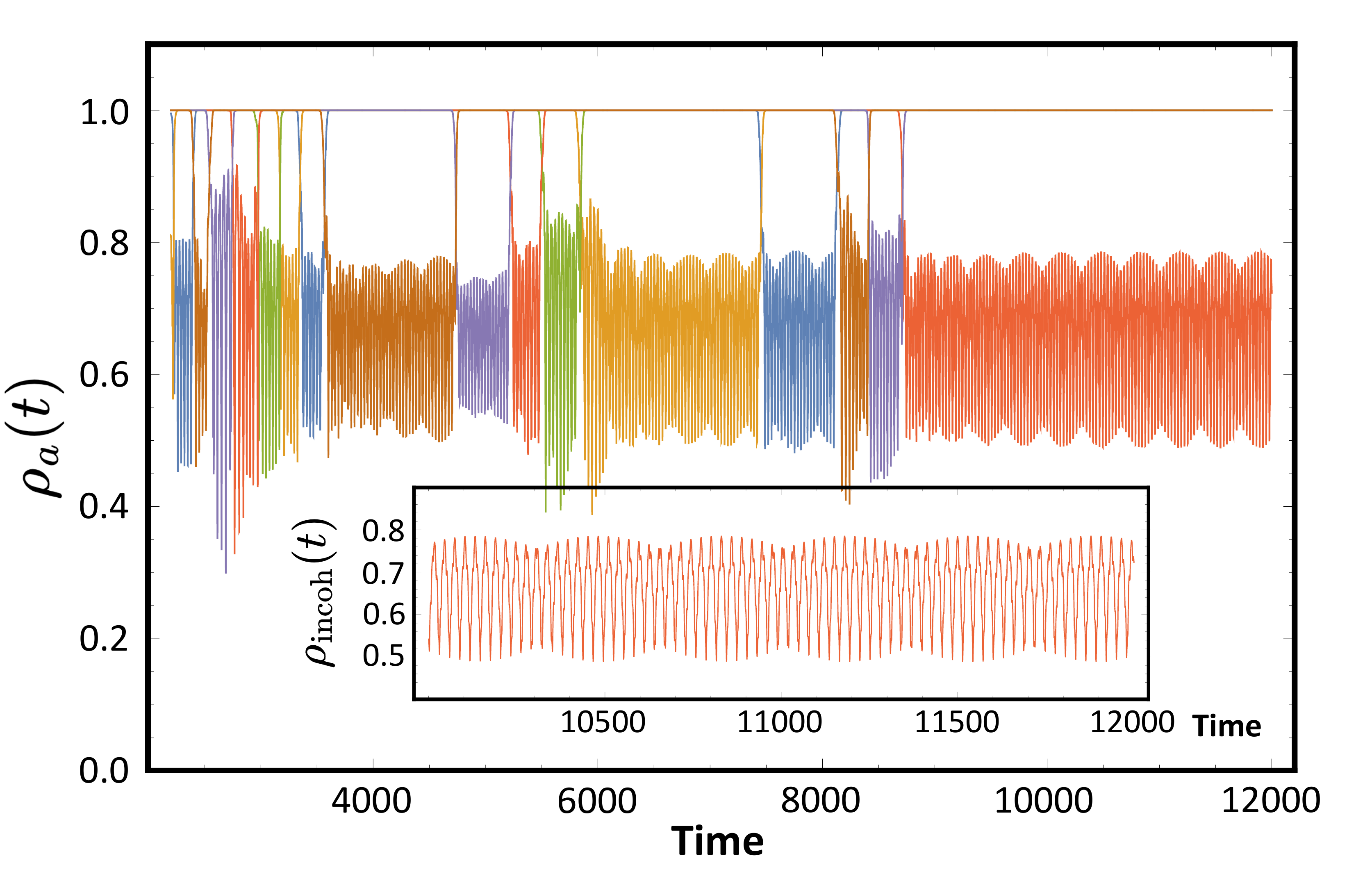}
\caption{Switching dynamics between quasiperiodic chimera states: Time evolution of the radial variables of the WS dynamics from a random initial condition with $A=0.5$ and $N=20$. The nonuniform constants of motion were obtained for $p=0.85$.} 
\label{Fig:WS_quasi}
\end{figure}

In contrast to the OA dynamics which is restricted to an invariant manifold where the phases are distributed according to the normalized Poisson kernel, taking nonuniform constants of motion in the WS transformation we can approach the dynamics outside the OA manifold. The nonuniform constants of motion are generated from~\cite{pikovsky_WS2} $\psi_j^{(a)} = (1-p)\frac{\pi}{2} + \frac{\pi p (j-1)}{N/2}$ and $\psi_{j+N/2}^{(a)} = -(1+p)\frac{\pi}{2} + \frac{\pi p (j-1)}{N/2}$ with $p=0.85$. A dynamical state which cannot be captured by the OA dynamics is the quasiperiodic chimera state reported in Ref.~\onlinecite{pikovsky_WS1}. In our network topology, we observe the switching dynamics between such quasiperiodic chimera states with nonuniform constants of motion. One realization for $p=0.85$, $A=0.5$ and $N=20$ is shown in Fig.~\ref{Fig:WS_quasi}. This observation underlines that the heteroclinic cycle is a robust phenomenon dictated by the symmetry of the network topology.

\section{\label{sec:nonidentical}Nonidentical oscillators}

In the following sections, we investigate a system of nonidentical oscillators in a ring of the six oscillator-populations. Here, we consider a heterogeneity characterized by $\gamma$ in Eqs.~(\ref{eq:OA_equation_radial}-\ref{eq:OA_equation_angular}) for the thermodynamic limit for which the OA manifold is known to be asymptotically attracting~\cite{attract_OA,laing_hetero1,OA2,lee1}. Furthermore, the natural frequency of the oscillator is generated from the Cauchy-Lorentz distribution for finite-sized systems in Eq.~(\ref{eq:micro_eq}).

\subsection{\label{subsec:small_hetero}Small heterogeneity: $\gamma=10^{-6}$}

\begin{figure}[t!]
\includegraphics[width=0.5\linewidth]{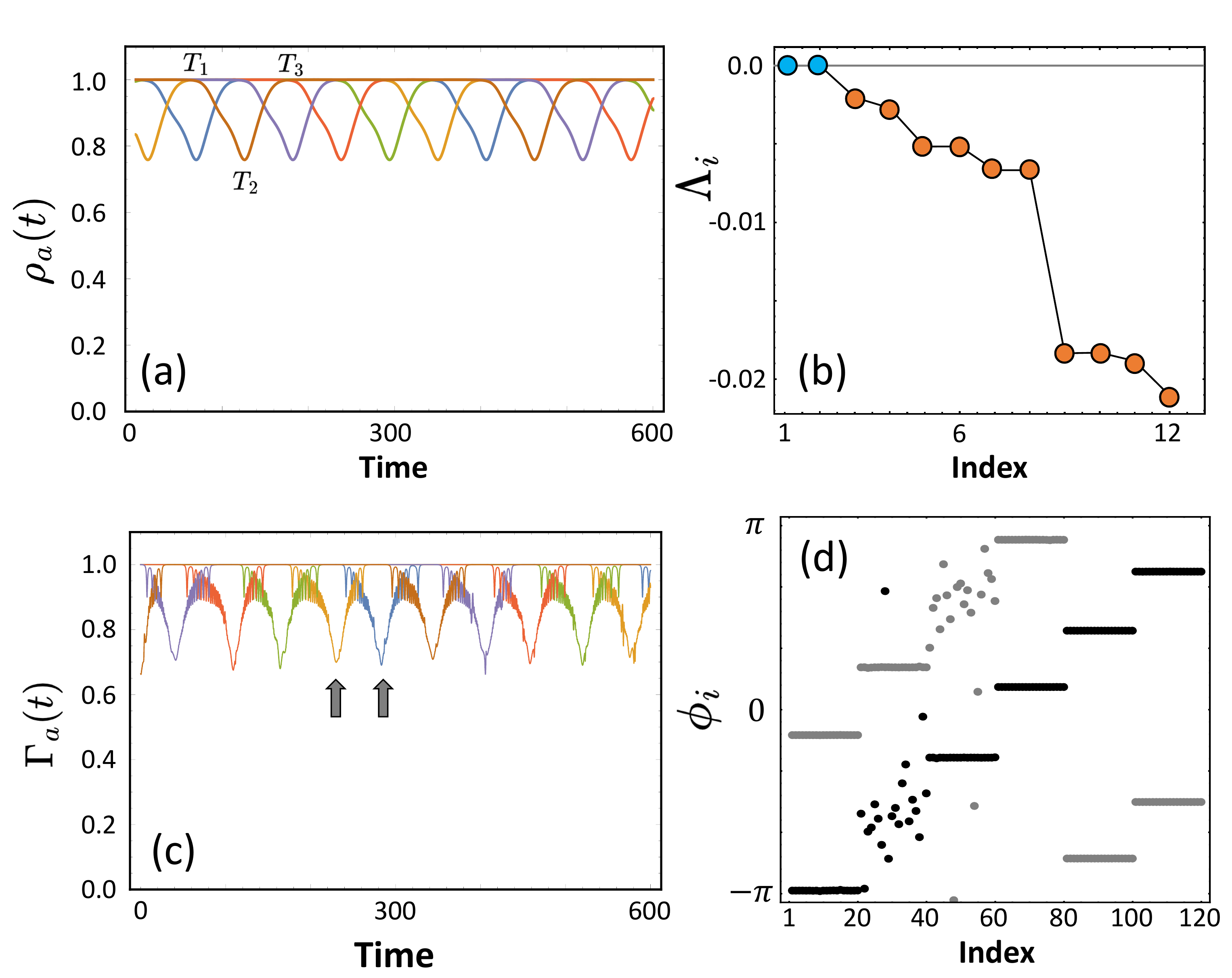}
\caption{(a) Time evolution of the radial variables of the OA dynamics after discarding transient behavior from a random initial condition. (b) Lyapunov exponents of the switching dynamics calculated along the switching trajectory. (c) Time evolution of moduli of the Kuramoto order parameters obtained from the microscopic dynamics with a random initial condition and $N=20$. The two arrows indicate the instants in time at which the snapshot in (d) were taken. (d) Phase snapshots of the microscopic dynamics for $N=20$ at two points indicated in (c). Other parameters: $\gamma=10^{-6}$ and $A=0.3$.} 
\label{fig:small_heterogeneity}
\end{figure}

First, we impose a small heterogeneity characterized by $\gamma = 10^{-6}$ on the natural frequencies of the oscillators. This small heterogeneity renders the Ott-Antonsen dynamics, which is known to be neutrally stable for strictly identical oscillators, attracting~\cite{attract_OA,laing_hetero1,OA2,lee1}. In Fig.~\ref{fig:small_heterogeneity} (a), time series of the radial variables of the slightly heterogeneous systems are depicted for $\gamma = 10^{-6}$ from a random initial condition. First of all, here we observe persistent switching of chimera states, which for identical oscillators was only detected in the presence of a low noise level. Yet, the switching phenomenology appears to be somewhat different. For most of time, the chimera state is characterized by four S-populations and two D-populations~\cite{comment1}. Consider, e.g., the evolution of the `brown' population in the time interval between $T_1$ and $T_3$ in Fig.~\ref{fig:small_heterogeneity} (a). In the first half of this time interval, i.e. up to $T_2$, it switches roles with the `blue' neighboring population, which becomes an S-population while the brown one becomes a D-population. As soon as the `blue' population has reached the S-state, the switching process with the other `purple' neighbor sets in. Consequently, the trajectory corresponds to a strict $\text{D}\text{S}^5$ chimera state or its symmetric counterparts only at periodic instants in time rather than during some time intervals. Furthermore, this persistent switching dynamics is attracting. This conjecture is confirmed by Lyapunov analysis. In Fig.~\ref{fig:small_heterogeneity} (b), Lyapunov exponents that were obtained along a switching trajectory are shown. All the LEs are negative except for two zero arising from time and phase shift invariance.

In order to study finite-sized ensembles, we need to directly investigate the microscopic dynamics in Eq.~(\ref{eq:micro_eq}) (note that the Watanabe-Strogatz ansatz does not work for heterogeneous oscillator ensembles~\cite{pikovsky_WS2}). First, we obtain the natural frequencies from the Cauchy-Lorentz distribution according to
\begin{flalign}
\frac{j-\frac{1}{2}}{N} &=\int_{-\infty}^{\omega_{j}}  g(\omega) d\omega = \frac{1}{2}+\frac{1}{\pi}\textrm{tan}^{-1} \big( \frac{\tilde{\omega}_j}{\gamma}\big)
\label{Eq:hetero-distribution}
\end{flalign}
for $j=1,...,N$, which produces $\{\omega_j = \gamma \textrm{tan}\big( \frac{\pi(2j-1-N)}{2N} \big) \}_{j=1}^{N}$. Directly solving the microscopic dynamics in Eq.~(\ref{eq:micro_eq}) with $\gamma=10^{-6}$, we observe a switching dynamics of the moduli of the Kuramoto order parameters defined in Eq.~(\ref{eq:Kuramoto_order_parameter}). 
In Fig.~\ref{fig:small_heterogeneity} (c), time evolution of the moduli of the Kuramoto order parameters is depicted. The envelop of them follows the same switching dynamics as in Fig.~\ref{fig:small_heterogeneity} (a), however with fluctuations superimposed, which stems from the finite-size effect. Two snapshots of the microscopic phases from a random initial condition are depicted in Fig.~\ref{fig:small_heterogeneity} (d), which are indicated by the two arrows in Fig.~\ref{fig:small_heterogeneity} (a). These are exactly the points in time at which there are five S- and one D-populations, corresponding to $\text{D}\text{S}^5$. As the $\text{D}\text{S}^5$ chimera state of the identical oscillator ensembles, the mean phases of the populations evolve as a nearly twisted state.

\subsection{\label{subsec:large_hetero}Larger heterogeneity: $\gamma=10^{-4}$}

\subsubsection{\label{subsec:hetero_stationary}Stationary Chimera States}

\begin{figure}[t!]
\includegraphics[width=0.5\linewidth]{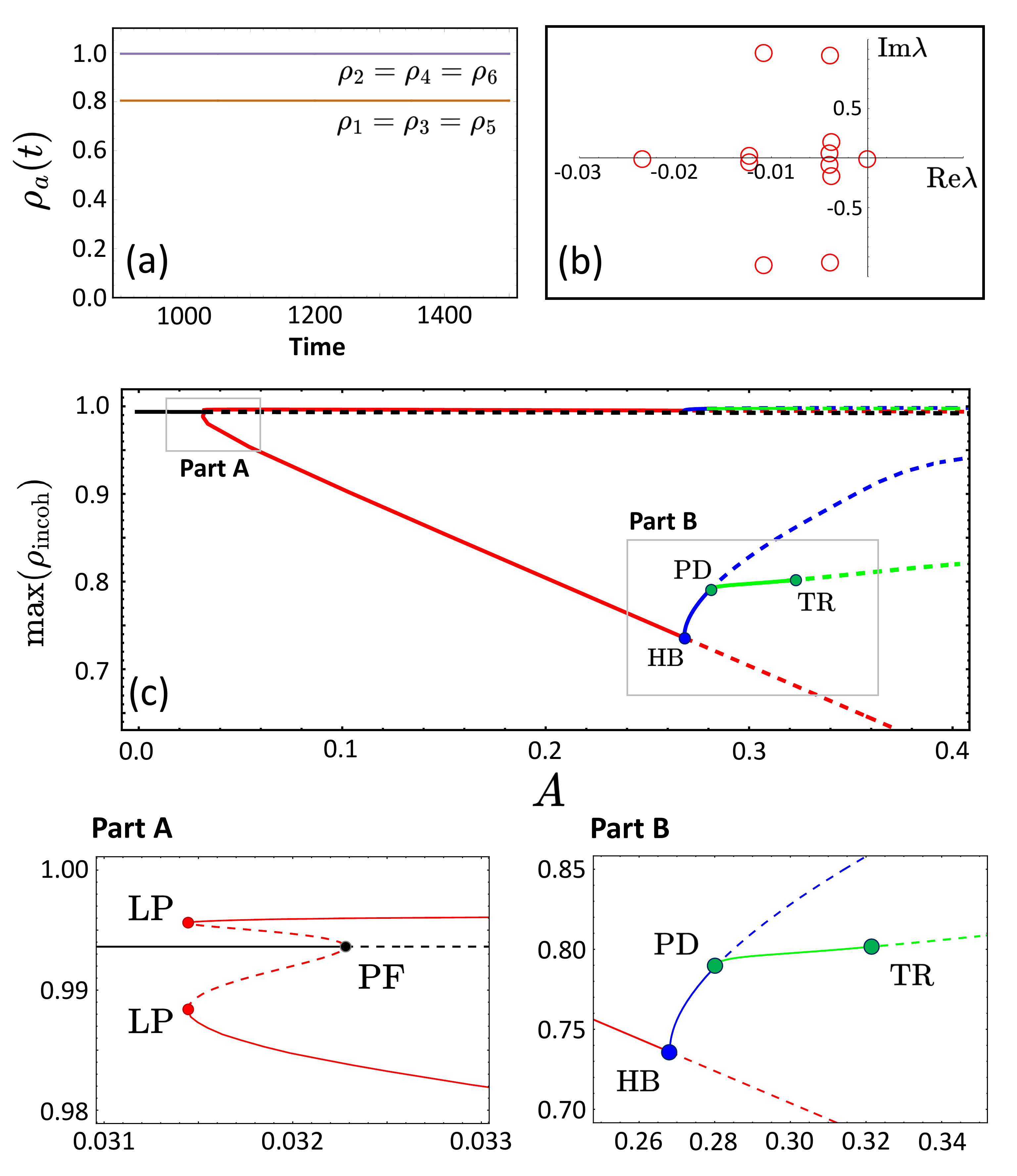}
\caption{(a) Time evolution of the radial variables for $(\text{D}\text{S})^3$ chimeras at $A=0.2$. (b) Eigenvalues of the Jacobian matrix evaluated at $(\text{D}\text{S})^3$ in the complex plane at $A=0
2$. (c) Bifurcation diagram of the $(\text{D}\text{S})^3$ chimera state. Solid and dashed lines indicate stable and unstable states, respectively. Black, red, blue and green: uniform states, stationary, breathing and period-doubled chimera solutions, respectively.} 
\label{Fig:hetero_stationary}
\end{figure}

Here, we consider a somewhat larger heterogeneity characterized by $\gamma=0.0001$ in Eqs.~(\ref{eq:OA_equation_radial}-\ref{eq:OA_equation_angular}). In this case, we do not observe any switching dynamics between chimera states from the numerical integration. We find an attracting stationary chimera state rather than the heteroclinic orbits between saddle chimeras. The attracting chimera state is of the type $(\text{D}\text{S})^3$. No other type of chimeras is observed within our numerical integration of Eqs.~(\ref{eq:OA_equation_radial}-\ref{eq:OA_equation_angular}) from random initial conditions. Looking at the network symmetry, the cluster $C_1=\{1,3,5\}$ and $C_2 = \{2,4,6\}$ are in fact intertwined clusters~\cite{pecora1,yscho2}, also called ISC set (independently synchronizable cluster set)~\cite{yscho,github}. This means that the stability of each cluster depends on the stability of the other cluster.

In Fig.~\ref{Fig:hetero_stationary} (a), time series of the radial variables of a stationary $(\text{D}\text{S})^3$ chimera starting from a random initial condition is shown for $A=0.2$. It is characterized by $\rho_a(t)=\rho_\text{D} <1$ (incoherent populations) for $a=1,3,5$ and $\rho_b(t) = \rho_\text{S} \approx 1$ (nearly-synchronized populations) for $b=2,4,6$. The phase variables are also locked at the common frequency. The phase differences between oscillators in the same cluster are found to be $\frac{2\pi}{3}$, i.e., $\varphi_{a+2}-\varphi_a = \frac{2\pi}{3}$. In Fig.~\ref{Fig:hetero_stationary} (b), the eigenvalues of the Jacobian matrix evaluated at $(\text{D}\text{S})^3$ are depicted in the complex plane. All the eigenvalues have negative real parts except for one zero corresponding to the phase shift invariance, confirming that the $(\text{D}\text{S})^3$ chimera is a linearly stable state.

\begin{figure}[t!]
\includegraphics[width=0.5\linewidth]{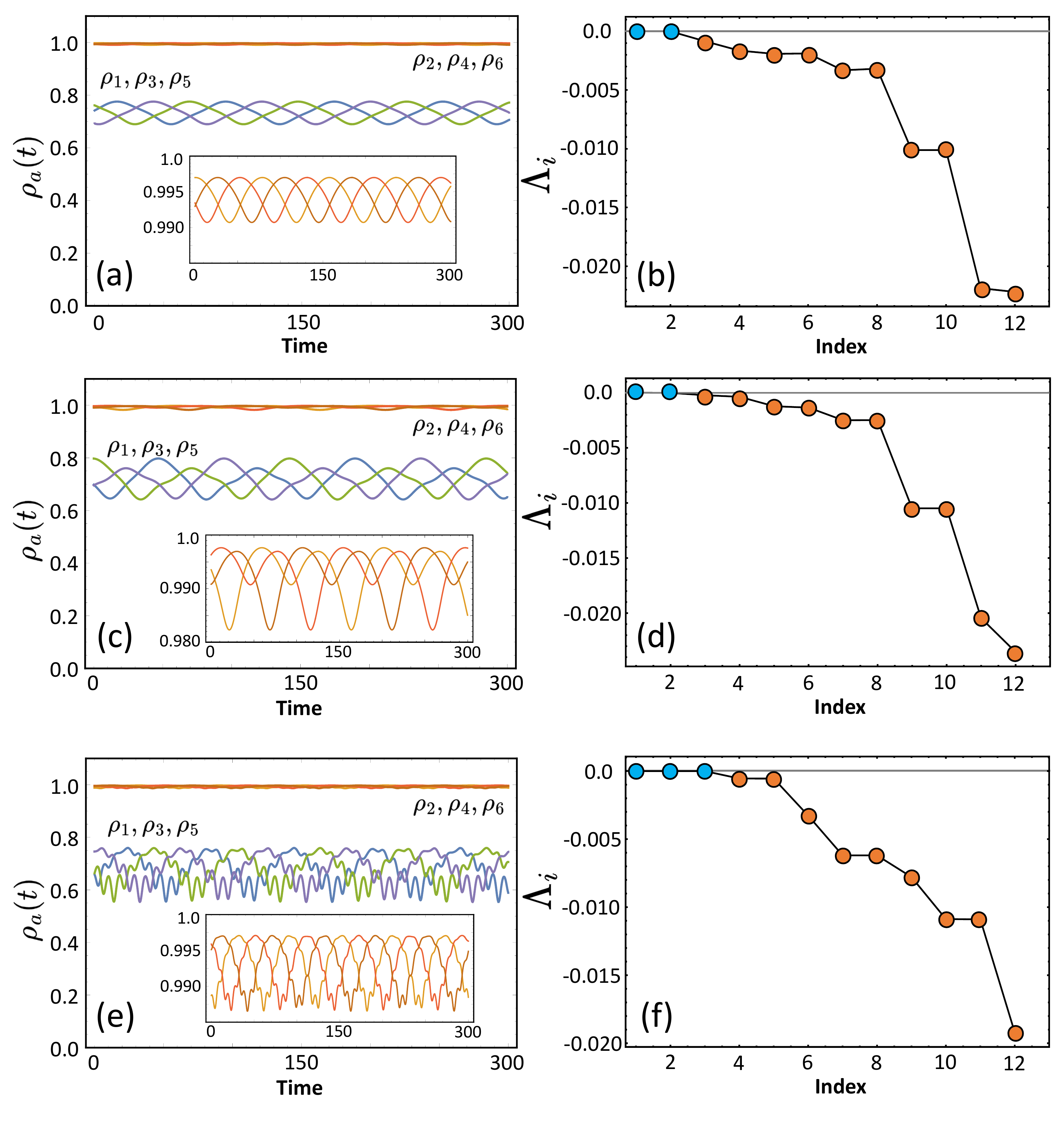}
\caption{ Breathing (a,b), period-doubled (c,d) and quasiperiodic (e,f) chimera states for $A=0.274$, $A=0.3$ and $A=0.34$, respectively. Left column: time series of the radial variables with insets of magnifying the dynamics of nearly-synchronized populations. Right column: Lyapunov exponents with the same color scheme as in Fig.~\ref{Fig:breathing_chimera}.} 
\label{Fig:hetero_breathing}
\end{figure}

In Fig.~\ref{Fig:hetero_stationary} (c), a bifurcation diagram of the $(\text{D}\text{S})^3$ chimera states is shown. For a small value of $A$, we find that a stable uniform solution (black) exists with $\rho_a = \rho_0 < 1$ for $a=1,...,6$ and equally spaced phase variables. We can interpret this uniform state as consisting of two clusters, $C_1$ and $C_2$ that are identical in their radial variables while the phase variables follow a twisted behavior. This uniform state is destabilized in a pitchfork bifurcation (PF) at $A_\text{PF}=0.0323$ in which one eigenvalue becomes positive. The eigenvector corresponding to the positive eigenvalue has the structure $\mathbf{v} = (\delta_+,\delta_{-},\delta_+,\delta_{-},\delta_+,\delta_{-})^\top$ where $\delta_+ \delta_{-}<0$. This means the uniform state is unstable along the transverse direction between two clusters. Due to the transversal instability, two symmetric solutions bifurcate from the uniform state in the pitchfork bifurcation, which is subcritical in our case (see Part A in Fig.~\ref{Fig:hetero_stationary}). Each of the two solutions possesses a nearly synchronized cluster and an incoherent cluster which form together a chimera of the type $(\text{D}\text{S})^3$ or $(\text{S}\text{D})^3$, respectively. Coming from low values of $A$, each of these two solution branches is born in a saddle-node bifurcation (LP) at $A_\text{LP} = 0.03144$ together with a stable, stationary $(\text{S}\text{D})^3$- respectively $(\text{D}\text{S})^3$-chimera state. This chimera state is stable in a wide range of the parameter $A$ as shown in Fig.~\ref{Fig:hetero_stationary} (c).

\subsubsection{\label{subsec:hetero_breathing}Breathing, Period-doubled and Quasiperiodic Chimera States}

The stationary $(\text{D}\text{S})^3$-type chimera states are destabilized in a supercritical Hopf bifurcation (HB) at $A_{\text{HB}}=0.26812$ giving rise to stable breathing chimera states. Trajectories of the radial variables are shown in Fig.~\ref{Fig:hetero_breathing} (a) together with the Lyapunov spectrum in Fig.~\ref{Fig:hetero_breathing} (b) for $A=0.274$. The oscillations of the three populations within one cluster are time-shifted by $T/3$ and $2T/3$ where $T$ is the period of the radial variable, as can be clearly seen for the incoherent population: $\rho_a(t) = \rho_{a+2}(t-\frac{T}{3}) = \rho_{a+4}(t-\frac{2T}{3})$ for $a=1,...,6$ (indices are taken modulo 6). 
Besides the two zero Lyapunov exponents, connected to the phase and time invariance, all Lyapunov exponents are negative, confirming that the breathing $(\text{D}\text{S})^3$-type chimera is also attracting. This state exists only in a small interval of the parameter $A$. At $A_{\text{PD}}=0.27995$, it is destabilized in a supercritical period-doubling bifurcation (PD), see Part B of Fig.~\ref{Fig:hetero_stationary} (c). Panels (c-d) in Fig.~\ref{Fig:hetero_breathing} confirm the period-doubled characteristics of the radial variables as well as the stability of the period-doubled chimera trajectory. Again, the radial variables exhibit the above specified spatiotemporal symmetry, with the difference that the period is nearly twice the period in Fig.~\ref{Fig:hetero_breathing} (a).

\begin{figure}[t!]
\includegraphics[width=0.5\linewidth]{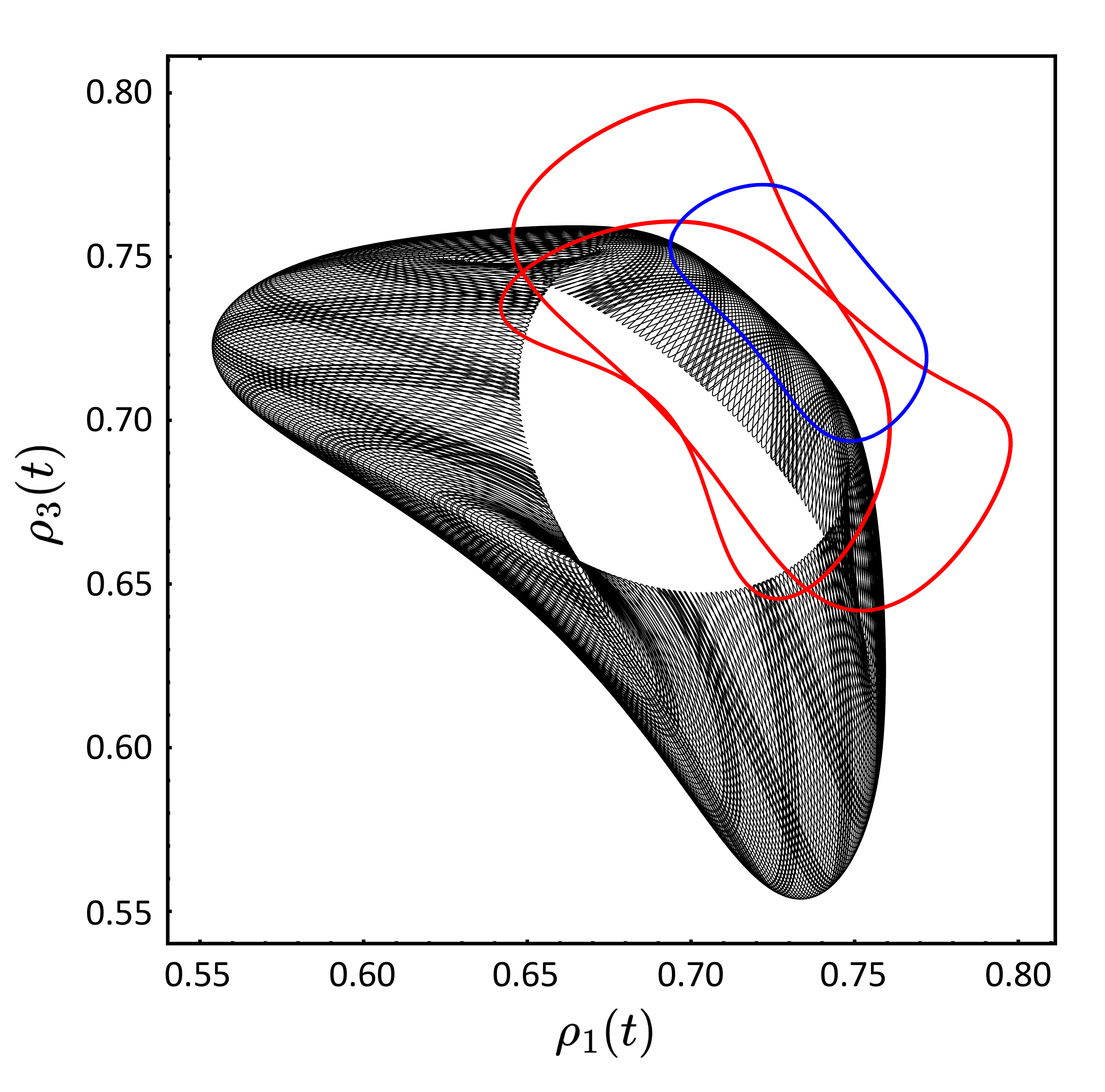}
\caption{Time-parametric plot of $\rho_1(t)$ vs. $\rho_3(t)$ for the time interval $\Delta t = 3000$ after discarding the transient behavior: breathing (blue), period-doubled (red) and quasiperiodic (black) chimera trajectories for $A=0.274$, $A=0.3$ and $A=0.34$, respectively.} 
\label{Fig:hetero_quasiperiodic}
\end{figure}

The period-doubled chimera state loses its stability in a supercritical torus bifurcation (TR) at $A_\text{TR}=0.31957$. For $A > A_\text{TR}$, one can observe a quasiperiodic chimera dynamics on a torus. An example dynamics and its stability are shown in Fig.~\ref{Fig:hetero_breathing} (e-f). The quasiperiodic chimera is characterized by one more zero Lyapunov exponent arising from the second incommensurate frequency. The rich dynamics of the $(\text{D}\text{S})^3$-type chimera states can be better appreciated in a time-parametric plot where $\rho_1(t)$ is plotted vs. $\rho_3(t)$ (Fig.~\ref{Fig:hetero_quasiperiodic}). The breathing chimera state appears as a simple, closed loop (blue) whereas the period-doubled chimera state (red) follows a double-wound loop in the projected space. In contrast, the spatiotemporal symmetry of the quasiperiodic motion on the torus is broken (black).

\section{\label{sec:conclusion}Conclusion and Outlook}

In this paper, we studied a system of six populations of identical Kuramoto-Sakaguchi phase oscillators in a ring topology. In the thermodynamic limit, the Ott-Antonsen dynamics possesses a variety of chimera solutions arising from the symmetry of the ring, most of them being unstable in nearly the entire parameter space. Only in a narrow interval of the inter-population coupling strength $\nu = 1-A$, we obtain stable chimera states characterized by one incoherent and five synchronized populations. These symmetric states are destabilized in a transcritical bifurcation and live in a large parameter region as saddle chimera states with one unstable direction. 
The six one-dimensional unstable manifolds of the saddle chimeras connect them in a heteroclinic cycle, rendering the chimera states observable in numerical integration from random initial conditions. Moreover, the trajectories display transient switching between the saddle chimera states, which becomes persistent when a small noise is imposed on the radial dynamics. At some large value of $A$, the stationary saddle chimeras undergo a Hopf bifurcation resulting in a heteroclinic orbit of saddle limit-cycles. Thus, we also observe switching between breathing chimeras. Moreover, in finite-sized populations, we observed in addition a heteroclinic switching between quasiperiodic chimera states. A heteroclinic cycle even persists in the presence of small heterogeneity in the oscillator natural frequencies. This robust occurrence of heteroclinic cycles with three different variants of base states (stationary, breathing and quasiperiodic) is in stark contrast to the dynamics of other network topologies. When using 3, 4, 5, 7, and 8 populations in a ring, we did not observe switching dynamics between chimera states from random initial conditions. Currently, we do not have a satisfactory explanation on why six populations behave differently.  However, we note that in three-population networks, switching chimeras were obtained with different coupling functions, more precisely, pairwise intra-population coupling 
with higher harmonics and sinusoidal nonpairwise inter-population coupling~\cite{bick_2018,Bick2019_m1}. It thus remains an interesting problem for future studies to investigate under which conditions observable heteroclinic cycles in oscillator networks exist.

Finally, we considered oscillator ensembles with a wider distribution of the heterogeneous natural frequencies in the thermodynamic limit. Here, instead of saddle chimera states, attracting chimera states with various symmetries and complex order parameter dynamics, dominate in phase space over a wide range of the parameter. The various macroscopic dynamics emerge in Hopf, period-doubling, and torus bifurcations.

In conclusion, we have discovered several types of heteroclinic switching observable in the macroscopic dynamics of nearly identical oscillator populations arranged in a ring. One might be tempted to interpret this as a further step towards understanding the mechanism of the dynamics of neural oscillator networks~\cite{Ashwin2016,tognoli2014metastable}, in particular, for encoding sequential information~\cite{bick_2018,Komarov_2009,skardal2020memory}. However, as our study on a strong heterogeneity revealed, one has to be very careful before drawing this conclusion for a biological system.

\section*{Data Availability Statement}
The data that support the findings of this study are available from the corresponding author upon reasonable request.

\begin{acknowledgments}
The authors would like to thank Young Sul Cho for providing additional computing facilities. This work has been supported by the Deutsche Forschungsgemeinschaft (project KR1189/18-2).
\end{acknowledgments}

\vskip 1cm 

\bibliography{aipchaos}

\providecommand{\noopsort}[1]{}\providecommand{\singleletter}[1]{#1}%
\begin{thebibliography}{63}%
\makeatletter
\providecommand \@ifxundefined [1]{%
 \@ifx{#1\undefined}
}%
\providecommand \@ifnum [1]{%
 \ifnum #1\expandafter \@firstoftwo
 \else \expandafter \@secondoftwo
 \fi
}%
\providecommand \@ifx [1]{%
 \ifx #1\expandafter \@firstoftwo
 \else \expandafter \@secondoftwo
 \fi
}%
\providecommand \natexlab [1]{#1}%
\providecommand \enquote  [1]{``#1''}%
\providecommand \bibnamefont  [1]{#1}%
\providecommand \bibfnamefont [1]{#1}%
\providecommand \citenamefont [1]{#1}%
\providecommand \href@noop [0]{\@secondoftwo}%
\providecommand \href [0]{\begingroup \@sanitize@url \@href}%
\providecommand \@href[1]{\@@startlink{#1}\@@href}%
\providecommand \@@href[1]{\endgroup#1\@@endlink}%
\providecommand \@sanitize@url [0]{\catcode `\\12\catcode `\$12\catcode
  `\&12\catcode `\#12\catcode `\^12\catcode `\_12\catcode `\%12\relax}%
\providecommand \@@startlink[1]{}%
\providecommand \@@endlink[0]{}%
\providecommand \url  [0]{\begingroup\@sanitize@url \@url }%
\providecommand \@url [1]{\endgroup\@href {#1}{\urlprefix }}%
\providecommand \urlprefix  [0]{URL }%
\providecommand \Eprint [0]{\href }%
\providecommand \doibase [0]{http://dx.doi.org/}%
\providecommand \selectlanguage [0]{\@gobble}%
\providecommand \bibinfo  [0]{\@secondoftwo}%
\providecommand \bibfield  [0]{\@secondoftwo}%
\providecommand \translation [1]{[#1]}%
\providecommand \BibitemOpen [0]{}%
\providecommand \bibitemStop [0]{}%
\providecommand \bibitemNoStop [0]{.\EOS\space}%
\providecommand \EOS [0]{\spacefactor3000\relax}%
\providecommand \BibitemShut  [1]{\csname bibitem#1\endcsname}%
\let\auto@bib@innerbib\@empty
\bibitem [{\citenamefont {Strogatz}(2003)}]{strogatz_sync}%
  \BibitemOpen
  \bibfield  {author} {\bibinfo {author} {\bibfnamefont {S.~H.}\ \bibnamefont
  {Strogatz}},\ }\href@noop {} {\emph {\bibinfo {title} {Sync}}}\ (\bibinfo
  {publisher} {Hyperion},\ \bibinfo {address} {New York},\ \bibinfo {year}
  {2003})\BibitemShut {NoStop}%
\bibitem [{\citenamefont {Pikovsky}, \citenamefont {Rosenblum},\ and\
  \citenamefont {Kurths}(2001)}]{pikovksy_sync}%
  \BibitemOpen
  \bibfield  {author} {\bibinfo {author} {\bibfnamefont {A.}~\bibnamefont
  {Pikovsky}}, \bibinfo {author} {\bibfnamefont {M.}~\bibnamefont {Rosenblum}},
  \ and\ \bibinfo {author} {\bibfnamefont {J.}~\bibnamefont {Kurths}},\
  }\href@noop {} {\emph {\bibinfo {title} {Synchronization: A Universal Concept
  in Nonlinear Sciences}}}\ (\bibinfo  {publisher} {Cambridge University
  Press},\ \bibinfo {address} {Cambridge},\ \bibinfo {year} {2001})\BibitemShut
  {NoStop}%
\bibitem [{chi(2022)}]{chimera22}%
  \BibitemOpen
  \bibfield  {title} {\enquote {\bibinfo {title} {Chimera states: From theory
  and experiments to technology and living systems},}\ \ }(\bibinfo
  {publisher} {MPIPKS, Dresden, Germany},\ \bibinfo {year} {2022})\ \bibinfo
  {note} {\url{https://www.pks.mpg.de/de/chimer22}}\BibitemShut {NoStop}%
\bibitem [{\citenamefont {Kuramoto}\ and\ \citenamefont
  {Battogtokh}(2002)}]{kuramoto2002}%
  \BibitemOpen
  \bibfield  {author} {\bibinfo {author} {\bibfnamefont {Y.}~\bibnamefont
  {Kuramoto}}\ and\ \bibinfo {author} {\bibfnamefont {D.}~\bibnamefont
  {Battogtokh}},\ }\bibfield  {title} {\enquote {\bibinfo {title} {Coexistence
  of coherence and incoherence in nonlocally coupled phase oscillators},}\
  }\href@noop {} {\bibfield  {journal} {\bibinfo  {journal} {Nonlinear Phenom.
  Complex Syst.}\ }\textbf {\bibinfo {volume} {5}},\ \bibinfo {pages} {380}
  (\bibinfo {year} {2002})}\BibitemShut {NoStop}%
\bibitem [{\citenamefont {Abrams}\ and\ \citenamefont
  {Strogatz}(2004)}]{abrams2004}%
  \BibitemOpen
  \bibfield  {author} {\bibinfo {author} {\bibfnamefont {D.~M.}\ \bibnamefont
  {Abrams}}\ and\ \bibinfo {author} {\bibfnamefont {S.~H.}\ \bibnamefont
  {Strogatz}},\ }\bibfield  {title} {\enquote {\bibinfo {title} {{Chimera
  States for Coupled Oscillators}},}\ }\href {\doibase
  10.1103/PhysRevLett.93.174102} {\bibfield  {journal} {\bibinfo  {journal}
  {Phys. Rev. Lett.}\ }\textbf {\bibinfo {volume} {93}},\ \bibinfo {pages}
  {174102} (\bibinfo {year} {2004})}\BibitemShut {NoStop}%
\bibitem [{\citenamefont {Panaggio}\ and\ \citenamefont
  {Abrams}(2015)}]{Panaggio_2015}%
  \BibitemOpen
  \bibfield  {author} {\bibinfo {author} {\bibfnamefont {M.~J.}\ \bibnamefont
  {Panaggio}}\ and\ \bibinfo {author} {\bibfnamefont {D.~M.}\ \bibnamefont
  {Abrams}},\ }\bibfield  {title} {\enquote {\bibinfo {title} {Chimera states:
  coexistence of coherence and incoherence in networks of coupled
  oscillators},}\ }\href {\doibase 10.1088/0951-7715/28/3/r67} {\bibfield
  {journal} {\bibinfo  {journal} {Nonlinearity}\ }\textbf {\bibinfo {volume}
  {28}},\ \bibinfo {pages} {R67--R87} (\bibinfo {year} {2015})}\BibitemShut
  {NoStop}%
\bibitem [{\citenamefont {Omel'chenko}(2018)}]{Omelchenko_2018}%
  \BibitemOpen
  \bibfield  {author} {\bibinfo {author} {\bibfnamefont {O.~E.}\ \bibnamefont
  {Omel'chenko}},\ }\bibfield  {title} {\enquote {\bibinfo {title} {The
  mathematics behind chimera states},}\ }\href {\doibase
  10.1088/1361-6544/aaaa07} {\bibfield  {journal} {\bibinfo  {journal}
  {Nonlinearity}\ }\textbf {\bibinfo {volume} {31}},\ \bibinfo {pages}
  {R121--R164} (\bibinfo {year} {2018})}\BibitemShut {NoStop}%
\bibitem [{\citenamefont {Montbri\'o}, \citenamefont {Kurths},\ and\
  \citenamefont {Blasius}(2004)}]{Kurths_twogroup}%
  \BibitemOpen
  \bibfield  {author} {\bibinfo {author} {\bibfnamefont {E.}~\bibnamefont
  {Montbri\'o}}, \bibinfo {author} {\bibfnamefont {J.}~\bibnamefont {Kurths}},
  \ and\ \bibinfo {author} {\bibfnamefont {B.}~\bibnamefont {Blasius}},\
  }\bibfield  {title} {\enquote {\bibinfo {title} {Synchronization of two
  interacting populations of oscillators},}\ }\href {\doibase
  10.1103/PhysRevE.70.056125} {\bibfield  {journal} {\bibinfo  {journal} {Phys.
  Rev. E}\ }\textbf {\bibinfo {volume} {70}},\ \bibinfo {pages} {056125}
  (\bibinfo {year} {2004})}\BibitemShut {NoStop}%
\bibitem [{\citenamefont {Abrams}\ \emph {et~al.}(2008)\citenamefont {Abrams},
  \citenamefont {Mirollo}, \citenamefont {Strogatz},\ and\ \citenamefont
  {Wiley}}]{abrams_chimera2008}%
  \BibitemOpen
  \bibfield  {author} {\bibinfo {author} {\bibfnamefont {D.~M.}\ \bibnamefont
  {Abrams}}, \bibinfo {author} {\bibfnamefont {R.}~\bibnamefont {Mirollo}},
  \bibinfo {author} {\bibfnamefont {S.~H.}\ \bibnamefont {Strogatz}}, \ and\
  \bibinfo {author} {\bibfnamefont {D.~A.}\ \bibnamefont {Wiley}},\ }\bibfield
  {title} {\enquote {\bibinfo {title} {Solvable model for chimera states of
  coupled oscillators},}\ }\href {\doibase 10.1103/PhysRevLett.101.084103}
  {\bibfield  {journal} {\bibinfo  {journal} {Phys. Rev. Lett.}\ }\textbf
  {\bibinfo {volume} {101}},\ \bibinfo {pages} {084103} (\bibinfo {year}
  {2008})}\BibitemShut {NoStop}%
\bibitem [{\citenamefont {Panaggio}\ \emph {et~al.}(2016)\citenamefont
  {Panaggio}, \citenamefont {Abrams}, \citenamefont {Ashwin},\ and\
  \citenamefont {Laing}}]{abrams_chimera2016}%
  \BibitemOpen
  \bibfield  {author} {\bibinfo {author} {\bibfnamefont {M.~J.}\ \bibnamefont
  {Panaggio}}, \bibinfo {author} {\bibfnamefont {D.~M.}\ \bibnamefont
  {Abrams}}, \bibinfo {author} {\bibfnamefont {P.}~\bibnamefont {Ashwin}}, \
  and\ \bibinfo {author} {\bibfnamefont {C.~R.}\ \bibnamefont {Laing}},\
  }\bibfield  {title} {\enquote {\bibinfo {title} {Chimera states in networks
  of phase oscillators: The case of two small populations},}\ }\href {\doibase
  10.1103/PhysRevE.93.012218} {\bibfield  {journal} {\bibinfo  {journal} {Phys.
  Rev. E}\ }\textbf {\bibinfo {volume} {93}},\ \bibinfo {pages} {012218}
  (\bibinfo {year} {2016})}\BibitemShut {NoStop}%
\bibitem [{\citenamefont {Lee}\ and\ \citenamefont {Krischer}(2021)}]{lee1}%
  \BibitemOpen
  \bibfield  {author} {\bibinfo {author} {\bibfnamefont {S.}~\bibnamefont
  {Lee}}\ and\ \bibinfo {author} {\bibfnamefont {K.}~\bibnamefont {Krischer}},\
  }\bibfield  {title} {\enquote {\bibinfo {title} {{Attracting Poisson chimeras
  in two-population networks}},}\ }\href {\doibase 10.1063/5.0065710}
  {\bibfield  {journal} {\bibinfo  {journal} {Chaos}\ }\textbf {\bibinfo
  {volume} {31}},\ \bibinfo {pages} {113101} (\bibinfo {year}
  {2021})}\BibitemShut {NoStop}%
\bibitem [{\citenamefont {Burylko}, \citenamefont {Martens},\ and\
  \citenamefont {Bick}(2022)}]{sym_twogroup}%
  \BibitemOpen
  \bibfield  {author} {\bibinfo {author} {\bibfnamefont {O.}~\bibnamefont
  {Burylko}}, \bibinfo {author} {\bibfnamefont {E.~A.}\ \bibnamefont
  {Martens}}, \ and\ \bibinfo {author} {\bibfnamefont {C.}~\bibnamefont
  {Bick}},\ }\bibfield  {title} {\enquote {\bibinfo {title} {Symmetry breaking
  yields chimeras in two small populations of kuramoto-type oscillators},}\
  }\href {\doibase 10.1063/5.0088465} {\bibfield  {journal} {\bibinfo
  {journal} {Chaos}\ }\textbf {\bibinfo {volume} {32}},\ \bibinfo {pages}
  {093109} (\bibinfo {year} {2022})}\BibitemShut {NoStop}%
\bibitem [{\citenamefont {Laing}(2019)}]{Laing_SL2019}%
  \BibitemOpen
  \bibfield  {author} {\bibinfo {author} {\bibfnamefont {C.~R.}\ \bibnamefont
  {Laing}},\ }\bibfield  {title} {\enquote {\bibinfo {title} {Dynamics and
  stability of chimera states in two coupled populations of oscillators},}\
  }\href {\doibase 10.1103/PhysRevE.100.042211} {\bibfield  {journal} {\bibinfo
   {journal} {Phys. Rev. E}\ }\textbf {\bibinfo {volume} {100}},\ \bibinfo
  {pages} {042211} (\bibinfo {year} {2019})}\BibitemShut {NoStop}%
\bibitem [{\citenamefont {Martens}(2010{\natexlab{a}})}]{martens_three}%
  \BibitemOpen
  \bibfield  {author} {\bibinfo {author} {\bibfnamefont {E.~A.}\ \bibnamefont
  {Martens}},\ }\bibfield  {title} {\enquote {\bibinfo {title} {Bistable
  chimera attractors on a triangular network of oscillator populations},}\
  }\href {\doibase 10.1103/PhysRevE.82.016216} {\bibfield  {journal} {\bibinfo
  {journal} {Phys. Rev. E}\ }\textbf {\bibinfo {volume} {82}},\ \bibinfo
  {pages} {016216} (\bibinfo {year} {2010}{\natexlab{a}})}\BibitemShut
  {NoStop}%
\bibitem [{\citenamefont {Martens}(2010{\natexlab{b}})}]{martens_three2}%
  \BibitemOpen
  \bibfield  {author} {\bibinfo {author} {\bibfnamefont {E.~A.}\ \bibnamefont
  {Martens}},\ }\bibfield  {title} {\enquote {\bibinfo {title} {Chimeras in a
  network of three oscillator populations with varying network topology},}\
  }\href {\doibase 10.1063/1.3499502} {\bibfield  {journal} {\bibinfo
  {journal} {Chaos}\ }\textbf {\bibinfo {volume} {20}},\ \bibinfo {pages}
  {043122} (\bibinfo {year} {2010}{\natexlab{b}})}\BibitemShut {NoStop}%
\bibitem [{\citenamefont {Laing}(2023)}]{laing_ring}%
  \BibitemOpen
  \bibfield  {author} {\bibinfo {author} {\bibfnamefont {C.~R.}\ \bibnamefont
  {Laing}},\ }\bibfield  {title} {\enquote {\bibinfo {title} {Chimeras on a
  ring of oscillator populations},}\ }\href {\doibase 10.1063/5.0127306}
  {\bibfield  {journal} {\bibinfo  {journal} {Chaos: An Interdisciplinary
  Journal of Nonlinear Science}\ }\textbf {\bibinfo {volume} {33}},\ \bibinfo
  {pages} {013121} (\bibinfo {year} {2023})}\BibitemShut {NoStop}%
\bibitem [{\citenamefont {Hong}, \citenamefont {Jo},\ and\ \citenamefont
  {Sin}(2013)}]{PhysRevE.88.032711}%
  \BibitemOpen
  \bibfield  {author} {\bibinfo {author} {\bibfnamefont {H.}~\bibnamefont
  {Hong}}, \bibinfo {author} {\bibfnamefont {J.}~\bibnamefont {Jo}}, \ and\
  \bibinfo {author} {\bibfnamefont {S.-J.}\ \bibnamefont {Sin}},\ }\bibfield
  {title} {\enquote {\bibinfo {title} {Stable and flexible system for glucose
  homeostasis},}\ }\href {\doibase 10.1103/PhysRevE.88.032711} {\bibfield
  {journal} {\bibinfo  {journal} {Phys. Rev. E}\ }\textbf {\bibinfo {volume}
  {88}},\ \bibinfo {pages} {032711} (\bibinfo {year} {2013})}\BibitemShut
  {NoStop}%
\bibitem [{\citenamefont {Pikovsky}\ and\ \citenamefont
  {Rosenblum}(2008)}]{pikovsky_WS1}%
  \BibitemOpen
  \bibfield  {author} {\bibinfo {author} {\bibfnamefont {A.}~\bibnamefont
  {Pikovsky}}\ and\ \bibinfo {author} {\bibfnamefont {M.}~\bibnamefont
  {Rosenblum}},\ }\bibfield  {title} {\enquote {\bibinfo {title} {Partially
  integrable dynamics of hierarchical populations of coupled oscillators},}\
  }\href {\doibase 10.1103/PhysRevLett.101.264103} {\bibfield  {journal}
  {\bibinfo  {journal} {Phys. Rev. Lett.}\ }\textbf {\bibinfo {volume} {101}},\
  \bibinfo {pages} {264103} (\bibinfo {year} {2008})}\BibitemShut {NoStop}%
\bibitem [{\citenamefont {Pikovsky}\ and\ \citenamefont
  {Rosenblum}(2011)}]{pikovsky_WS2}%
  \BibitemOpen
  \bibfield  {author} {\bibinfo {author} {\bibfnamefont {A.}~\bibnamefont
  {Pikovsky}}\ and\ \bibinfo {author} {\bibfnamefont {M.}~\bibnamefont
  {Rosenblum}},\ }\bibfield  {title} {\enquote {\bibinfo {title} {Dynamics of
  heterogeneous oscillator ensembles in terms of collective variables},}\
  }\href {\doibase https://doi.org/10.1016/j.physd.2011.01.002} {\bibfield
  {journal} {\bibinfo  {journal} {Physica D: Nonlinear Phenomena}\ }\textbf
  {\bibinfo {volume} {240}},\ \bibinfo {pages} {872--881} (\bibinfo {year}
  {2011})}\BibitemShut {NoStop}%
\bibitem [{\citenamefont {Martens}, \citenamefont {Bick},\ and\ \citenamefont
  {Panaggio}(2016)}]{hetero_twogroup}%
  \BibitemOpen
  \bibfield  {author} {\bibinfo {author} {\bibfnamefont {E.~A.}\ \bibnamefont
  {Martens}}, \bibinfo {author} {\bibfnamefont {C.}~\bibnamefont {Bick}}, \
  and\ \bibinfo {author} {\bibfnamefont {M.~J.}\ \bibnamefont {Panaggio}},\
  }\bibfield  {title} {\enquote {\bibinfo {title} {Chimera states in two
  populations with heterogeneous phase-lag},}\ }\href {\doibase
  10.1063/1.4958930} {\bibfield  {journal} {\bibinfo  {journal} {Chaos}\
  }\textbf {\bibinfo {volume} {26}},\ \bibinfo {pages} {094819} (\bibinfo
  {year} {2016})}\BibitemShut {NoStop}%
\bibitem [{\citenamefont {Paz\'o}\ and\ \citenamefont
  {Montbri\'o}(2014)}]{pazo_winfree}%
  \BibitemOpen
  \bibfield  {author} {\bibinfo {author} {\bibfnamefont {D.}~\bibnamefont
  {Paz\'o}}\ and\ \bibinfo {author} {\bibfnamefont {E.}~\bibnamefont
  {Montbri\'o}},\ }\bibfield  {title} {\enquote {\bibinfo {title}
  {Low-dimensional dynamics of populations of pulse-coupled oscillators},}\
  }\href {\doibase 10.1103/PhysRevX.4.011009} {\bibfield  {journal} {\bibinfo
  {journal} {Phys. Rev. X}\ }\textbf {\bibinfo {volume} {4}},\ \bibinfo {pages}
  {011009} (\bibinfo {year} {2014})}\BibitemShut {NoStop}%
\bibitem [{\citenamefont {Olmi}(2015)}]{olmi_chaos}%
  \BibitemOpen
  \bibfield  {author} {\bibinfo {author} {\bibfnamefont {S.}~\bibnamefont
  {Olmi}},\ }\bibfield  {title} {\enquote {\bibinfo {title} {Chimera states in
  coupled kuramoto oscillators with inertia},}\ }\href {\doibase
  10.1063/1.4938734} {\bibfield  {journal} {\bibinfo  {journal} {Chaos}\
  }\textbf {\bibinfo {volume} {25}},\ \bibinfo {pages} {123125} (\bibinfo
  {year} {2015})}\BibitemShut {NoStop}%
\bibitem [{\citenamefont {Olmi}\ \emph {et~al.}(2015)\citenamefont {Olmi},
  \citenamefont {Martens}, \citenamefont {Thutupalli},\ and\ \citenamefont
  {Torcini}}]{Olmi_rotator}%
  \BibitemOpen
  \bibfield  {author} {\bibinfo {author} {\bibfnamefont {S.}~\bibnamefont
  {Olmi}}, \bibinfo {author} {\bibfnamefont {E.~A.}\ \bibnamefont {Martens}},
  \bibinfo {author} {\bibfnamefont {S.}~\bibnamefont {Thutupalli}}, \ and\
  \bibinfo {author} {\bibfnamefont {A.}~\bibnamefont {Torcini}},\ }\bibfield
  {title} {\enquote {\bibinfo {title} {Intermittent chaotic chimeras for
  coupled rotators},}\ }\href {\doibase 10.1103/PhysRevE.92.030901} {\bibfield
  {journal} {\bibinfo  {journal} {Phys. Rev. E}\ }\textbf {\bibinfo {volume}
  {92}},\ \bibinfo {pages} {030901} (\bibinfo {year} {2015})}\BibitemShut
  {NoStop}%
\bibitem [{\citenamefont {Bick}\ and\ \citenamefont
  {Ashwin}(2016)}]{Bick_2016_chaotic}%
  \BibitemOpen
  \bibfield  {author} {\bibinfo {author} {\bibfnamefont {C.}~\bibnamefont
  {Bick}}\ and\ \bibinfo {author} {\bibfnamefont {P.}~\bibnamefont {Ashwin}},\
  }\bibfield  {title} {\enquote {\bibinfo {title} {Chaotic weak chimeras and
  their persistence in coupled populations of phase oscillators},}\ }\href
  {\doibase 10.1088/0951-7715/29/5/1468} {\bibfield  {journal} {\bibinfo
  {journal} {Nonlinearity}\ }\textbf {\bibinfo {volume} {29}},\ \bibinfo
  {pages} {1468} (\bibinfo {year} {2016})}\BibitemShut {NoStop}%
\bibitem [{\citenamefont {Semenova}\ \emph {et~al.}(2016)\citenamefont
  {Semenova}, \citenamefont {Zakharova}, \citenamefont {Anishchenko},\ and\
  \citenamefont {Sch\"oll}}]{alternating1}%
  \BibitemOpen
  \bibfield  {author} {\bibinfo {author} {\bibfnamefont {N.}~\bibnamefont
  {Semenova}}, \bibinfo {author} {\bibfnamefont {A.}~\bibnamefont {Zakharova}},
  \bibinfo {author} {\bibfnamefont {V.}~\bibnamefont {Anishchenko}}, \ and\
  \bibinfo {author} {\bibfnamefont {E.}~\bibnamefont {Sch\"oll}},\ }\bibfield
  {title} {\enquote {\bibinfo {title} {Coherence-resonance chimeras in a
  network of excitable elements},}\ }\href {\doibase
  10.1103/PhysRevLett.117.014102} {\bibfield  {journal} {\bibinfo  {journal}
  {Phys. Rev. Lett.}\ }\textbf {\bibinfo {volume} {117}},\ \bibinfo {pages}
  {014102} (\bibinfo {year} {2016})}\BibitemShut {NoStop}%
\bibitem [{\citenamefont {Buscarino}\ \emph {et~al.}(2015)\citenamefont
  {Buscarino}, \citenamefont {Frasca}, \citenamefont {Gambuzza},\ and\
  \citenamefont {H\"ovel}}]{timevarying_twogroup}%
  \BibitemOpen
  \bibfield  {author} {\bibinfo {author} {\bibfnamefont {A.}~\bibnamefont
  {Buscarino}}, \bibinfo {author} {\bibfnamefont {M.}~\bibnamefont {Frasca}},
  \bibinfo {author} {\bibfnamefont {L.~V.}\ \bibnamefont {Gambuzza}}, \ and\
  \bibinfo {author} {\bibfnamefont {P.}~\bibnamefont {H\"ovel}},\ }\bibfield
  {title} {\enquote {\bibinfo {title} {Chimera states in time-varying complex
  networks},}\ }\href {\doibase 10.1103/PhysRevE.91.022817} {\bibfield
  {journal} {\bibinfo  {journal} {Phys. Rev. E}\ }\textbf {\bibinfo {volume}
  {91}},\ \bibinfo {pages} {022817} (\bibinfo {year} {2015})}\BibitemShut
  {NoStop}%
\bibitem [{\citenamefont {Laing}(2012)}]{alternating2}%
  \BibitemOpen
  \bibfield  {author} {\bibinfo {author} {\bibfnamefont {C.~R.}\ \bibnamefont
  {Laing}},\ }\bibfield  {title} {\enquote {\bibinfo {title} {Disorder-induced
  dynamics in a pair of coupled heterogeneous phase oscillator networks},}\
  }\href {\doibase 10.1063/1.4758814} {\bibfield  {journal} {\bibinfo
  {journal} {Chaos: An Interdisciplinary Journal of Nonlinear Science}\
  }\textbf {\bibinfo {volume} {22}},\ \bibinfo {pages} {043104} (\bibinfo
  {year} {2012})}\BibitemShut {NoStop}%
\bibitem [{\citenamefont {Ma}, \citenamefont {Wang},\ and\ \citenamefont
  {Liu}(2010)}]{ma2010robust}%
  \BibitemOpen
  \bibfield  {author} {\bibinfo {author} {\bibfnamefont {R.}~\bibnamefont
  {Ma}}, \bibinfo {author} {\bibfnamefont {J.}~\bibnamefont {Wang}}, \ and\
  \bibinfo {author} {\bibfnamefont {Z.}~\bibnamefont {Liu}},\ }\bibfield
  {title} {\enquote {\bibinfo {title} {Robust features of chimera states and
  the implementation of alternating chimera states},}\ }\href@noop {}
  {\bibfield  {journal} {\bibinfo  {journal} {Europhysics Letters}\ }\textbf
  {\bibinfo {volume} {91}},\ \bibinfo {pages} {40006} (\bibinfo {year}
  {2010})}\BibitemShut {NoStop}%
\bibitem [{\citenamefont {Zhang}\ \emph {et~al.}(2020)\citenamefont {Zhang},
  \citenamefont {Nicolaou}, \citenamefont {Hart}, \citenamefont {Roy},\ and\
  \citenamefont {Motter}}]{motter_switching}%
  \BibitemOpen
  \bibfield  {author} {\bibinfo {author} {\bibfnamefont {Y.}~\bibnamefont
  {Zhang}}, \bibinfo {author} {\bibfnamefont {Z.~G.}\ \bibnamefont {Nicolaou}},
  \bibinfo {author} {\bibfnamefont {J.~D.}\ \bibnamefont {Hart}}, \bibinfo
  {author} {\bibfnamefont {R.}~\bibnamefont {Roy}}, \ and\ \bibinfo {author}
  {\bibfnamefont {A.~E.}\ \bibnamefont {Motter}},\ }\bibfield  {title}
  {\enquote {\bibinfo {title} {Critical switching in globally attractive
  chimeras},}\ }\href {\doibase 10.1103/PhysRevX.10.011044} {\bibfield
  {journal} {\bibinfo  {journal} {Phys. Rev. X}\ }\textbf {\bibinfo {volume}
  {10}},\ \bibinfo {pages} {011044} (\bibinfo {year} {2020})}\BibitemShut
  {NoStop}%
\bibitem [{\citenamefont {Bick}(2018)}]{bick_2018}%
  \BibitemOpen
  \bibfield  {author} {\bibinfo {author} {\bibfnamefont {C.}~\bibnamefont
  {Bick}},\ }\bibfield  {title} {\enquote {\bibinfo {title} {Heteroclinic
  switching between chimeras},}\ }\href {\doibase 10.1103/PhysRevE.97.050201}
  {\bibfield  {journal} {\bibinfo  {journal} {Phys. Rev. E}\ }\textbf {\bibinfo
  {volume} {97}},\ \bibinfo {pages} {050201} (\bibinfo {year}
  {2018})}\BibitemShut {NoStop}%
\bibitem [{\citenamefont {Bick}(2019)}]{Bick2019_m1}%
  \BibitemOpen
  \bibfield  {author} {\bibinfo {author} {\bibfnamefont {C.}~\bibnamefont
  {Bick}},\ }\bibfield  {title} {\enquote {\bibinfo {title} {Heteroclinic
  dynamics of localized frequency synchrony: Heteroclinic cycles for small
  populations},}\ }\href {\doibase 10.1007/s00332-019-09552-5} {\bibfield
  {journal} {\bibinfo  {journal} {Journal of Nonlinear Science}\ }\textbf
  {\bibinfo {volume} {29}},\ \bibinfo {pages} {2547--2570} (\bibinfo {year}
  {2019})}\BibitemShut {NoStop}%
\bibitem [{\citenamefont {Bick}\ and\ \citenamefont
  {Lohse}(2019)}]{Bick019_m2}%
  \BibitemOpen
  \bibfield  {author} {\bibinfo {author} {\bibfnamefont {C.}~\bibnamefont
  {Bick}}\ and\ \bibinfo {author} {\bibfnamefont {A.}~\bibnamefont {Lohse}},\
  }\bibfield  {title} {\enquote {\bibinfo {title} {Heteroclinic dynamics of
  localized frequency synchrony: Stability of heteroclinic cycles and
  networks},}\ }\href {\doibase 10.1007/s00332-019-09562-3} {\bibfield
  {journal} {\bibinfo  {journal} {Journal of Nonlinear Science}\ }\textbf
  {\bibinfo {volume} {29}},\ \bibinfo {pages} {2571--2600} (\bibinfo {year}
  {2019})}\BibitemShut {NoStop}%
\bibitem [{\citenamefont {Haugland}, \citenamefont {Schmidt},\ and\
  \citenamefont {Krischer}(2015)}]{Haugland2015}%
  \BibitemOpen
  \bibfield  {author} {\bibinfo {author} {\bibfnamefont {S.~W.}\ \bibnamefont
  {Haugland}}, \bibinfo {author} {\bibfnamefont {L.}~\bibnamefont {Schmidt}}, \
  and\ \bibinfo {author} {\bibfnamefont {K.}~\bibnamefont {Krischer}},\
  }\bibfield  {title} {\enquote {\bibinfo {title} {Self-organized alternating
  chimera states in oscillatory media},}\ }\href {\doibase 10.1038/srep09883}
  {\bibfield  {journal} {\bibinfo  {journal} {Scientific Reports}\ }\textbf
  {\bibinfo {volume} {5}},\ \bibinfo {pages} {9883} (\bibinfo {year}
  {2015})}\BibitemShut {NoStop}%
\bibitem [{\citenamefont {Goldschmidt}, \citenamefont {Pikovsky},\ and\
  \citenamefont {Politi}(2019)}]{blinking}%
  \BibitemOpen
  \bibfield  {author} {\bibinfo {author} {\bibfnamefont {R.~J.}\ \bibnamefont
  {Goldschmidt}}, \bibinfo {author} {\bibfnamefont {A.}~\bibnamefont
  {Pikovsky}}, \ and\ \bibinfo {author} {\bibfnamefont {A.}~\bibnamefont
  {Politi}},\ }\bibfield  {title} {\enquote {\bibinfo {title} {Blinking
  chimeras in globally coupled rotators},}\ }\href {\doibase 10.1063/1.5105367}
  {\bibfield  {journal} {\bibinfo  {journal} {Chaos: An Interdisciplinary
  Journal of Nonlinear Science}\ }\textbf {\bibinfo {volume} {29}},\ \bibinfo
  {pages} {071101} (\bibinfo {year} {2019})}\BibitemShut {NoStop}%
\bibitem [{\citenamefont {Ebrahimzadeh}\ \emph {et~al.}(2020)\citenamefont
  {Ebrahimzadeh}, \citenamefont {Schiek}, \citenamefont {Jaros}, \citenamefont
  {Kapitaniak}, \citenamefont {van Waasen},\ and\ \citenamefont
  {Maistrenko}}]{Ebrahimzadeh2020}%
  \BibitemOpen
  \bibfield  {author} {\bibinfo {author} {\bibfnamefont {P.}~\bibnamefont
  {Ebrahimzadeh}}, \bibinfo {author} {\bibfnamefont {M.}~\bibnamefont
  {Schiek}}, \bibinfo {author} {\bibfnamefont {P.}~\bibnamefont {Jaros}},
  \bibinfo {author} {\bibfnamefont {T.}~\bibnamefont {Kapitaniak}}, \bibinfo
  {author} {\bibfnamefont {S.}~\bibnamefont {van Waasen}}, \ and\ \bibinfo
  {author} {\bibfnamefont {Y.}~\bibnamefont {Maistrenko}},\ }\bibfield  {title}
  {\enquote {\bibinfo {title} {Minimal chimera states in phase-lag coupled
  mechanical oscillators},}\ }\href {\doibase 10.1140/epjst/e2020-900270-4}
  {\bibfield  {journal} {\bibinfo  {journal} {The European Physical Journal
  Special Topics}\ }\textbf {\bibinfo {volume} {229}},\ \bibinfo {pages}
  {2205--2214} (\bibinfo {year} {2020})}\BibitemShut {NoStop}%
\bibitem [{\citenamefont {Brezetsky}\ \emph {et~al.}(2021)\citenamefont
  {Brezetsky}, \citenamefont {Jaros}, \citenamefont {Levchenko}, \citenamefont
  {Kapitaniak},\ and\ \citenamefont {Maistrenko}}]{chimera_complex}%
  \BibitemOpen
  \bibfield  {author} {\bibinfo {author} {\bibfnamefont {S.}~\bibnamefont
  {Brezetsky}}, \bibinfo {author} {\bibfnamefont {P.}~\bibnamefont {Jaros}},
  \bibinfo {author} {\bibfnamefont {R.}~\bibnamefont {Levchenko}}, \bibinfo
  {author} {\bibfnamefont {T.}~\bibnamefont {Kapitaniak}}, \ and\ \bibinfo
  {author} {\bibfnamefont {Y.}~\bibnamefont {Maistrenko}},\ }\bibfield  {title}
  {\enquote {\bibinfo {title} {Chimera complexity},}\ }\href {\doibase
  10.1103/PhysRevE.103.L050204} {\bibfield  {journal} {\bibinfo  {journal}
  {Phys. Rev. E}\ }\textbf {\bibinfo {volume} {103}},\ \bibinfo {pages}
  {L050204} (\bibinfo {year} {2021})}\BibitemShut {NoStop}%
\bibitem [{\citenamefont {Ashwin}\ and\ \citenamefont
  {Burylko}(2015)}]{weak_chimera}%
  \BibitemOpen
  \bibfield  {author} {\bibinfo {author} {\bibfnamefont {P.}~\bibnamefont
  {Ashwin}}\ and\ \bibinfo {author} {\bibfnamefont {O.}~\bibnamefont
  {Burylko}},\ }\bibfield  {title} {\enquote {\bibinfo {title} {Weak chimeras
  in minimal networks of coupled phase oscillators},}\ }\href {\doibase
  10.1063/1.4905197} {\bibfield  {journal} {\bibinfo  {journal} {Chaos: An
  Interdisciplinary Journal of Nonlinear Science}\ }\textbf {\bibinfo {volume}
  {25}},\ \bibinfo {pages} {013106} (\bibinfo {year} {2015})}\BibitemShut
  {NoStop}%
\bibitem [{\citenamefont {Ott}\ and\ \citenamefont {Antonsen}(2008)}]{OA1}%
  \BibitemOpen
  \bibfield  {author} {\bibinfo {author} {\bibfnamefont {E.}~\bibnamefont
  {Ott}}\ and\ \bibinfo {author} {\bibfnamefont {T.~M.}\ \bibnamefont
  {Antonsen}},\ }\bibfield  {title} {\enquote {\bibinfo {title} {Low
  dimensional behavior of large systems of globally coupled oscillators},}\
  }\href {\doibase 10.1063/1.2930766} {\bibfield  {journal} {\bibinfo
  {journal} {Chaos}\ }\textbf {\bibinfo {volume} {18}},\ \bibinfo {pages}
  {037113} (\bibinfo {year} {2008})}\BibitemShut {NoStop}%
\bibitem [{\citenamefont {Ott}\ and\ \citenamefont {Antonsen}(2009)}]{OA2}%
  \BibitemOpen
  \bibfield  {author} {\bibinfo {author} {\bibfnamefont {E.}~\bibnamefont
  {Ott}}\ and\ \bibinfo {author} {\bibfnamefont {T.~M.}\ \bibnamefont
  {Antonsen}},\ }\bibfield  {title} {\enquote {\bibinfo {title} {Long time
  evolution of phase oscillator systems},}\ }\href {\doibase 10.1063/1.3136851}
  {\bibfield  {journal} {\bibinfo  {journal} {Chaos}\ }\textbf {\bibinfo
  {volume} {19}},\ \bibinfo {pages} {023117} (\bibinfo {year}
  {2009})}\BibitemShut {NoStop}%
\bibitem [{\citenamefont {Marvel}, \citenamefont {Mirollo},\ and\ \citenamefont
  {Strogatz}(2009)}]{WS_mobius}%
  \BibitemOpen
  \bibfield  {author} {\bibinfo {author} {\bibfnamefont {S.~A.}\ \bibnamefont
  {Marvel}}, \bibinfo {author} {\bibfnamefont {R.~E.}\ \bibnamefont {Mirollo}},
  \ and\ \bibinfo {author} {\bibfnamefont {S.~H.}\ \bibnamefont {Strogatz}},\
  }\bibfield  {title} {\enquote {\bibinfo {title} {Identical phase oscillators
  with global sinusoidal coupling evolve by möbius group action},}\ }\href
  {\doibase 10.1063/1.3247089} {\bibfield  {journal} {\bibinfo  {journal}
  {Chaos}\ }\textbf {\bibinfo {volume} {19}},\ \bibinfo {pages} {043104}
  (\bibinfo {year} {2009})}\BibitemShut {NoStop}%
\bibitem [{\citenamefont {Watanabe}\ and\ \citenamefont
  {Strogatz}(1994)}]{WS_original2}%
  \BibitemOpen
  \bibfield  {author} {\bibinfo {author} {\bibfnamefont {S.}~\bibnamefont
  {Watanabe}}\ and\ \bibinfo {author} {\bibfnamefont {S.~H.}\ \bibnamefont
  {Strogatz}},\ }\bibfield  {title} {\enquote {\bibinfo {title} {{Constants of
  motion for superconducting Josephson arrays}},}\ }\href {\doibase
  https://doi.org/10.1016/0167-2789(94)90196-1} {\bibfield  {journal} {\bibinfo
   {journal} {Physica D: Nonlinear Phenomena}\ }\textbf {\bibinfo {volume}
  {74}},\ \bibinfo {pages} {197--253} (\bibinfo {year} {1994})}\BibitemShut
  {NoStop}%
\bibitem [{\citenamefont {Bick}\ \emph {et~al.}(2020)\citenamefont {Bick},
  \citenamefont {Goodfellow}, \citenamefont {Laing},\ and\ \citenamefont
  {Martens}}]{Bick2020}%
  \BibitemOpen
  \bibfield  {author} {\bibinfo {author} {\bibfnamefont {C.}~\bibnamefont
  {Bick}}, \bibinfo {author} {\bibfnamefont {M.}~\bibnamefont {Goodfellow}},
  \bibinfo {author} {\bibfnamefont {C.~R.}\ \bibnamefont {Laing}}, \ and\
  \bibinfo {author} {\bibfnamefont {E.~A.}\ \bibnamefont {Martens}},\
  }\bibfield  {title} {\enquote {\bibinfo {title} {Understanding the dynamics
  of biological and neural oscillator networks through exact mean-field
  reductions: a review},}\ }\href {\doibase 10.1186/s13408-020-00086-9}
  {\bibfield  {journal} {\bibinfo  {journal} {The Journal of Mathematical
  Neuroscience}\ }\textbf {\bibinfo {volume} {10}},\ \bibinfo {pages} {9}
  (\bibinfo {year} {2020})}\BibitemShut {NoStop}%
\bibitem [{\citenamefont {Laing}(2009{\natexlab{a}})}]{Laing_OA}%
  \BibitemOpen
  \bibfield  {author} {\bibinfo {author} {\bibfnamefont {C.~R.}\ \bibnamefont
  {Laing}},\ }\bibfield  {title} {\enquote {\bibinfo {title} {The dynamics of
  chimera states in heterogeneous kuramoto networks},}\ }\href {\doibase
  https://doi.org/10.1016/j.physd.2009.04.012} {\bibfield  {journal} {\bibinfo
  {journal} {Physica D: Nonlinear Phenomena}\ }\textbf {\bibinfo {volume}
  {238}},\ \bibinfo {pages} {1569--1588} (\bibinfo {year}
  {2009}{\natexlab{a}})}\BibitemShut {NoStop}%
\bibitem [{\citenamefont {Lee}, \citenamefont {Cho},\ and\ \citenamefont
  {Hong}(2018)}]{lee_twisted}%
  \BibitemOpen
  \bibfield  {author} {\bibinfo {author} {\bibfnamefont {S.}~\bibnamefont
  {Lee}}, \bibinfo {author} {\bibfnamefont {Y.~S.}\ \bibnamefont {Cho}}, \ and\
  \bibinfo {author} {\bibfnamefont {H.}~\bibnamefont {Hong}},\ }\bibfield
  {title} {\enquote {\bibinfo {title} {Twisted states in low-dimensional
  hypercubic lattices},}\ }\href {\doibase 10.1103/PhysRevE.98.062221}
  {\bibfield  {journal} {\bibinfo  {journal} {Phys. Rev. E}\ }\textbf {\bibinfo
  {volume} {98}},\ \bibinfo {pages} {062221} (\bibinfo {year}
  {2018})}\BibitemShut {NoStop}%
\bibitem [{\citenamefont {Cho}, \citenamefont {Nishikawa},\ and\ \citenamefont
  {Motter}(2017)}]{yscho}%
  \BibitemOpen
  \bibfield  {author} {\bibinfo {author} {\bibfnamefont {Y.~S.}\ \bibnamefont
  {Cho}}, \bibinfo {author} {\bibfnamefont {T.}~\bibnamefont {Nishikawa}}, \
  and\ \bibinfo {author} {\bibfnamefont {A.~E.}\ \bibnamefont {Motter}},\
  }\bibfield  {title} {\enquote {\bibinfo {title} {Stable chimeras and
  independently synchronizable clusters},}\ }\href {\doibase
  10.1103/PhysRevLett.119.084101} {\bibfield  {journal} {\bibinfo  {journal}
  {Phys. Rev. Lett.}\ }\textbf {\bibinfo {volume} {119}},\ \bibinfo {pages}
  {084101} (\bibinfo {year} {2017})}\BibitemShut {NoStop}%
\bibitem [{\citenamefont {Pecora}\ \emph {et~al.}(2014)\citenamefont {Pecora},
  \citenamefont {Sorrentino}, \citenamefont {Hagerstrom}, \citenamefont
  {Murphy},\ and\ \citenamefont {Roy}}]{pecora1}%
  \BibitemOpen
  \bibfield  {author} {\bibinfo {author} {\bibfnamefont {L.~M.}\ \bibnamefont
  {Pecora}}, \bibinfo {author} {\bibfnamefont {F.}~\bibnamefont {Sorrentino}},
  \bibinfo {author} {\bibfnamefont {A.~M.}\ \bibnamefont {Hagerstrom}},
  \bibinfo {author} {\bibfnamefont {T.~E.}\ \bibnamefont {Murphy}}, \ and\
  \bibinfo {author} {\bibfnamefont {R.}~\bibnamefont {Roy}},\ }\bibfield
  {title} {\enquote {\bibinfo {title} {Cluster synchronization and isolated
  desynchronization in complex networks with symmetries},}\ }\href
  {{https://www.nature.com/articles/ncomms5079}} {\bibfield  {journal}
  {\bibinfo  {journal} {Nat. Commun.}\ }\textbf {\bibinfo {volume} {5}},\
  \bibinfo {pages} {4079} (\bibinfo {year} {2014})}\BibitemShut {NoStop}%
\bibitem [{\citenamefont {Sorrentino}\ \emph {et~al.}(2016)\citenamefont
  {Sorrentino}, \citenamefont {Pecora}, \citenamefont {Hagerstrom},
  \citenamefont {Murphy},\ and\ \citenamefont {Roy}}]{pecora2}%
  \BibitemOpen
  \bibfield  {author} {\bibinfo {author} {\bibfnamefont {F.}~\bibnamefont
  {Sorrentino}}, \bibinfo {author} {\bibfnamefont {L.}~\bibnamefont {Pecora}},
  \bibinfo {author} {\bibfnamefont {A.~M.}\ \bibnamefont {Hagerstrom}},
  \bibinfo {author} {\bibfnamefont {T.~E.}\ \bibnamefont {Murphy}}, \ and\
  \bibinfo {author} {\bibfnamefont {R.}~\bibnamefont {Roy}},\ }\bibfield
  {title} {\enquote {\bibinfo {title} {Complete characterization of the
  stability of cluster synchronization in complex dynamical networks},}\
  }\href@noop {} {\bibfield  {journal} {\bibinfo  {journal} {Sci. Adv.}\
  }\textbf {\bibinfo {volume} {2}},\ \bibinfo {pages} {e1501737} (\bibinfo
  {year} {2016})}\BibitemShut {NoStop}%
\bibitem [{\citenamefont {Inc.}()}]{Mathematica}%
  \BibitemOpen
  \bibfield  {author} {\bibinfo {author} {\bibfnamefont {W.~R.}\ \bibnamefont
  {Inc.}},\ }\href {https://www.wolfram.com/mathematica} {\enquote {\bibinfo
  {title} {Mathematica, {V}ersion 12.0},}\ }\bibinfo {note} {Champaign, IL,
  2022: Numerical integration was performed in \textit{NDSolve} with
  \textit{IDA}}\BibitemShut {NoStop}%
\bibitem [{\citenamefont {Lee}\ and\ \citenamefont {Krischer}(2022)}]{nts}%
  \BibitemOpen
  \bibfield  {author} {\bibinfo {author} {\bibfnamefont {S.}~\bibnamefont
  {Lee}}\ and\ \bibinfo {author} {\bibfnamefont {K.}~\bibnamefont {Krischer}},\
  }\bibfield  {title} {\enquote {\bibinfo {title} {{Nontrivial twisted states
  in nonlocally coupled Stuart-Landau oscillators}},}\ }\href {\doibase
  10.1103/PhysRevE.106.044210} {\bibfield  {journal} {\bibinfo  {journal}
  {Phys. Rev. E}\ }\textbf {\bibinfo {volume} {106}},\ \bibinfo {pages}
  {044210} (\bibinfo {year} {2022})}\BibitemShut {NoStop}%
\bibitem [{\citenamefont {Pikovsky}\ and\ \citenamefont
  {Politi}(2016)}]{pikovsky_LE}%
  \BibitemOpen
  \bibfield  {author} {\bibinfo {author} {\bibfnamefont {A.}~\bibnamefont
  {Pikovsky}}\ and\ \bibinfo {author} {\bibfnamefont {A.}~\bibnamefont
  {Politi}},\ }\href@noop {} {\emph {\bibinfo {title} {Lyapunov Exponents: A
  Tool to Explore Complex Dynamics}}}\ (\bibinfo  {publisher} {Cambridge
  University Press},\ \bibinfo {address} {Cambridge},\ \bibinfo {year}
  {2016})\BibitemShut {NoStop}%
\bibitem [{\citenamefont {Oseledets}(1968)}]{oseledets}%
  \BibitemOpen
  \bibfield  {author} {\bibinfo {author} {\bibfnamefont {V.}~\bibnamefont
  {Oseledets}},\ }\bibfield  {title} {\enquote {\bibinfo {title} {{A
  multiplicative ergodic theorem. Characteristic Liapunov, exponents of
  dynamical systems}},}\ }\href@noop {} {\bibfield  {journal} {\bibinfo
  {journal} {Trans. Mosc. Math. Soc.}\ }\textbf {\bibinfo {volume} {19}},\
  \bibinfo {pages} {197} (\bibinfo {year} {1968})}\BibitemShut {NoStop}%
\bibitem [{\citenamefont {Ginelli}\ \emph {et~al.}(2013)\citenamefont
  {Ginelli}, \citenamefont {Chat\'{e}}, \citenamefont {Livi},\ and\
  \citenamefont {Politi}}]{CLV1}%
  \BibitemOpen
  \bibfield  {author} {\bibinfo {author} {\bibfnamefont {F.}~\bibnamefont
  {Ginelli}}, \bibinfo {author} {\bibfnamefont {H.}~\bibnamefont {Chat\'{e}}},
  \bibinfo {author} {\bibfnamefont {R.}~\bibnamefont {Livi}}, \ and\ \bibinfo
  {author} {\bibfnamefont {A.}~\bibnamefont {Politi}},\ }\bibfield  {title}
  {\enquote {\bibinfo {title} {{Covariant Lyapunov vectors}},}\ }\href@noop {}
  {\bibfield  {journal} {\bibinfo  {journal} {J. Phys. A: Math. Theor.}\
  }\textbf {\bibinfo {volume} {46}},\ \bibinfo {pages} {254005} (\bibinfo
  {year} {2013})}\BibitemShut {NoStop}%
\bibitem [{\citenamefont {Kuptsov}\ and\ \citenamefont {Parlitz}(2012)}]{CLV2}%
  \BibitemOpen
  \bibfield  {author} {\bibinfo {author} {\bibfnamefont {P.~V.}\ \bibnamefont
  {Kuptsov}}\ and\ \bibinfo {author} {\bibfnamefont {U.}~\bibnamefont
  {Parlitz}},\ }\bibfield  {title} {\enquote {\bibinfo {title} {{Theory and
  computation of covariant Lyapunov vectors}},}\ }\href@noop {} {\bibfield
  {journal} {\bibinfo  {journal} {J. Nonlinear Sci.}\ }\textbf {\bibinfo
  {volume} {22}},\ \bibinfo {pages} {727} (\bibinfo {year} {2012})}\BibitemShut
  {NoStop}%
\bibitem [{\citenamefont {Lee}\ and\ \citenamefont {Krischer}(2023)}]{lee2}%
  \BibitemOpen
  \bibfield  {author} {\bibinfo {author} {\bibfnamefont {S.}~\bibnamefont
  {Lee}}\ and\ \bibinfo {author} {\bibfnamefont {K.}~\bibnamefont {Krischer}},\
  }\bibfield  {title} {\enquote {\bibinfo {title} {Chaotic chimera attractors
  in a triangular network of identical oscillators},}\ }\href {\doibase
  10.1103/PhysRevE.107.054205} {\bibfield  {journal} {\bibinfo  {journal}
  {Phys. Rev. E}\ }\textbf {\bibinfo {volume} {107}},\ \bibinfo {pages}
  {054205} (\bibinfo {year} {2023})}\BibitemShut {NoStop}%
\bibitem [{\citenamefont {Pietras}\ and\ \citenamefont
  {Daffertshofer}(2016)}]{attract_OA}%
  \BibitemOpen
  \bibfield  {author} {\bibinfo {author} {\bibfnamefont {B.}~\bibnamefont
  {Pietras}}\ and\ \bibinfo {author} {\bibfnamefont {A.}~\bibnamefont
  {Daffertshofer}},\ }\bibfield  {title} {\enquote {\bibinfo {title}
  {Ott-antonsen attractiveness for parameter-dependent oscillatory systems},}\
  }\href {\doibase 10.1063/1.4963371} {\bibfield  {journal} {\bibinfo
  {journal} {Chaos: An Interdisciplinary Journal of Nonlinear Science}\
  }\textbf {\bibinfo {volume} {26}},\ \bibinfo {pages} {103101} (\bibinfo
  {year} {2016})}\BibitemShut {NoStop}%
\bibitem [{\citenamefont {Laing}(2009{\natexlab{b}})}]{laing_hetero1}%
  \BibitemOpen
  \bibfield  {author} {\bibinfo {author} {\bibfnamefont {C.~R.}\ \bibnamefont
  {Laing}},\ }\bibfield  {title} {\enquote {\bibinfo {title} {Chimera states in
  heterogeneous networks},}\ }\href {\doibase 10.1063/1.3068353} {\bibfield
  {journal} {\bibinfo  {journal} {Chaos: An Interdisciplinary Journal of
  Nonlinear Science}\ }\textbf {\bibinfo {volume} {19}},\ \bibinfo {pages}
  {013113} (\bibinfo {year} {2009}{\natexlab{b}})}\BibitemShut {NoStop}%
\bibitem [{com()}]{comment1}%
  \BibitemOpen
  \bibinfo {note} {Note that the small heterogeneity prevents the full
  synchronization of a population. Yet, we keep the same notation and refer to
  nearly synchronized populations with $\rho_a \in [0.995,1)$ as
  $\text{S}$-populations.}\BibitemShut {Stop}%
\bibitem [{\citenamefont {Cho}(2019)}]{yscho2}%
  \BibitemOpen
  \bibfield  {author} {\bibinfo {author} {\bibfnamefont {Y.~S.}\ \bibnamefont
  {Cho}},\ }\bibfield  {title} {\enquote {\bibinfo {title} {Concurrent
  formation of nearly synchronous clusters in each intertwined cluster set with
  parameter mismatches},}\ }\href {\doibase 10.1103/PhysRevE.99.052215}
  {\bibfield  {journal} {\bibinfo  {journal} {Phys. Rev. E}\ }\textbf {\bibinfo
  {volume} {99}},\ \bibinfo {pages} {052215} (\bibinfo {year}
  {2019})}\BibitemShut {NoStop}%
\bibitem [{git()}]{github}%
  \BibitemOpen
  \bibinfo {note} {To find cluster patterns, see
  \url{https://github.com/tnishi0/grouping-clusters/}}\BibitemShut {NoStop}%
\bibitem [{\citenamefont {Ashwin}, \citenamefont {Coombes},\ and\ \citenamefont
  {Nicks}(2016)}]{Ashwin2016}%
  \BibitemOpen
  \bibfield  {author} {\bibinfo {author} {\bibfnamefont {P.}~\bibnamefont
  {Ashwin}}, \bibinfo {author} {\bibfnamefont {S.}~\bibnamefont {Coombes}}, \
  and\ \bibinfo {author} {\bibfnamefont {R.}~\bibnamefont {Nicks}},\ }\bibfield
   {title} {\enquote {\bibinfo {title} {Mathematical frameworks for oscillatory
  network dynamics in neuroscience},}\ }\href {\doibase
  10.1186/s13408-015-0033-6} {\bibfield  {journal} {\bibinfo  {journal} {The
  Journal of Mathematical Neuroscience}\ }\textbf {\bibinfo {volume} {6}},\
  \bibinfo {pages} {2} (\bibinfo {year} {2016})}\BibitemShut {NoStop}%
\bibitem [{\citenamefont {Tognoli}\ and\ \citenamefont
  {Kelso}(2014)}]{tognoli2014metastable}%
  \BibitemOpen
  \bibfield  {author} {\bibinfo {author} {\bibfnamefont {E.}~\bibnamefont
  {Tognoli}}\ and\ \bibinfo {author} {\bibfnamefont {J.~S.}\ \bibnamefont
  {Kelso}},\ }\bibfield  {title} {\enquote {\bibinfo {title} {The metastable
  brain},}\ }\href@noop {} {\bibfield  {journal} {\bibinfo  {journal} {Neuron}\
  }\textbf {\bibinfo {volume} {81}},\ \bibinfo {pages} {35--48} (\bibinfo
  {year} {2014})}\BibitemShut {NoStop}%
\bibitem [{\citenamefont {Komarov}, \citenamefont {Osipov},\ and\ \citenamefont
  {Suykens}(2009)}]{Komarov_2009}%
  \BibitemOpen
  \bibfield  {author} {\bibinfo {author} {\bibfnamefont {M.~A.}\ \bibnamefont
  {Komarov}}, \bibinfo {author} {\bibfnamefont {G.~V.}\ \bibnamefont {Osipov}},
  \ and\ \bibinfo {author} {\bibfnamefont {J.~A.~K.}\ \bibnamefont {Suykens}},\
  }\bibfield  {title} {\enquote {\bibinfo {title} {Sequentially activated
  groups in neural networks},}\ }\href {\doibase 10.1209/0295-5075/86/60006}
  {\bibfield  {journal} {\bibinfo  {journal} {Europhysics Letters}\ }\textbf
  {\bibinfo {volume} {86}},\ \bibinfo {pages} {60006} (\bibinfo {year}
  {2009})}\BibitemShut {NoStop}%
\bibitem [{\citenamefont {Skardal}\ and\ \citenamefont
  {Arenas}(2020)}]{skardal2020memory}%
  \BibitemOpen
  \bibfield  {author} {\bibinfo {author} {\bibfnamefont {P.~S.}\ \bibnamefont
  {Skardal}}\ and\ \bibinfo {author} {\bibfnamefont {A.}~\bibnamefont
  {Arenas}},\ }\bibfield  {title} {\enquote {\bibinfo {title} {{Memory
  selection and information switching in oscillator networks with higher-order
  interactions}},}\ }\href
  {{https://iopscience.iop.org/article/10.1088/2632-072X/abbd4c}} {\bibfield
  {journal} {\bibinfo  {journal} {Journal of Physics: Complexity}\ }\textbf
  {\bibinfo {volume} {2}},\ \bibinfo {pages} {015003} (\bibinfo {year}
  {2020})}\BibitemShut {NoStop}%
\end{thebibliography}%


\end{document}